\documentstyle[psfig]{mn}

\newif\ifAMStwofonts
\AMStwofontstrue

\newcommand{\be}{\begin{equation}}
\newcommand{\ee}{\end{equation}}
\newcommand{\ba}{\begin{eqnarray}}
\newcommand{\ea}{\end{eqnarray}}
\newcommand{\brr}{\begin{array}}
\newcommand{\err}{\end{array}}
\newcommand{\bc}{\begin{center}}
\newcommand{\ec}{\end{center}}
\newcommand{\hm}{\,h^{-1}{\rm Mpc}}

\newcommand{\lum}{\,{\rm erg\,s^{-1}}}

\newcommand{\mincir}{\raise
  -2.truept\hbox{\rlap{\hbox{$\sim$}}\raise5.truept \hbox{$<$}\ }}
\newcommand{\magcir}{\raise
  -2.truept\hbox{\rlap{\hbox{$\sim$}}\raise5.truept \hbox{$>$}\ }}
\newcommand{\siml}{\raise
  -2.truept\hbox{\rlap{\hbox{$\sim$}}\raise5.truept \hbox{$<$}\ }}
\newcommand{\simg}{\raise
  -2.truept\hbox{\rlap{\hbox{$\sim$}}\raise5.truept \hbox{$>$}\ }}

\ifoldfss
\ifCUPmtlplainloaded \else
\NewTextAlphabet{textbfit} {cmbxti10} {}
\NewTextAlphabet{textbfss} {cmssbx10} {}
\NewMathAlphabet{mathbfit} {cmbxti10} {} 
\NewMathAlphabet{mathbfss} {cmssbx10} {} 
\fi
\ifAMStwofonts
\ifCUPmtlplainloaded \else
\NewSymbolFont{upmath} {eurm10}
\NewSymbolFont{AMSa} {msam10}
\NewMathSymbol{\upi}     {0}{upmath}{19}
\NewMathSymbol{\umu}     {0}{upmath}{16}
\NewMathSymbol{\upartial}{0}{upmath}{40}
\NewMathSymbol{\leqslant}{3}{AMSa}{36}
\NewMathSymbol{\geqslant}{3}{AMSa}{3E}

\fi
\fi
\fi 

\ifnfssone
\newmathalphabet{\mathit}
\addtoversion{normal}{\mathit}{cmr}{m}{it}
\addtoversion{bold}{\mathit}{cmr}{bx}{it}
\newmathalphabet{\mathbfit} 
\addtoversion{normal}{\mathbfit}{cmr}{bx}{it}
\addtoversion{bold}{\mathbfit}{cmr}{bx}{it}
\newmathalphabet{\mathbfss} 
\addtoversion{normal}{\mathbfss}{cmss}{bx}{n}
\addtoversion{bold}{\mathbfss}{cmss}{bx}{n}
\ifAMStwofonts
\ifCUPmtlplainloaded \else
\UseAMStwoboldmath
\makeatletter
\new@mathgroup\upmath@group
\define@mathgroup\mv@normal\upmath@group{eur}{m}{n}
\define@mathgroup\mv@bold\upmath@group{eur}{b}{n}
\edef\UPM{\hexnumber\upmath@group}
\new@mathgroup\amsa@group
\define@mathgroup\mv@normal\amsa@group{msa}{m}{n}
\define@mathgroup\mv@bold\amsa@group{msa}{m}{n}
\edef\AMSa{\hexnumber\amsa@group}
\makeatother
\mathchardef\upi="0\UPM19
\mathchardef\umu="0\UPM16
\mathchardef\upartial="0\UPM40
\mathchardef\leqslant="3\AMSa36
\mathchardef\geqslant="3\AMSa3E
\fi
\fi
\fi 

\ifnfsstwo
\DeclareMathAlphabet{\mathbfit}{OT1}{cmr}{bx}{it}
\SetMathAlphabet\mathbfit{bold}{OT1}{cmr}{bx}{it}
\DeclareMathAlphabet{\mathbfss}{OT1}{cmss}{bx}{n}
\SetMathAlphabet\mathbfss{bold}{OT1}{cmss}{bx}{n}
\ifAMStwofonts
\ifCUPmtlplainloaded \else
\DeclareSymbolFont{UPM}{U}{eur}{m}{n}
\SetSymbolFont{UPM}{bold}{U}{eur}{b}{n}
\DeclareSymbolFont{AMSa}{U}{msa}{m}{n}
\DeclareMathSymbol{\upi}{0}{UPM}{"19}
\DeclareMathSymbol{\umu}{0}{UPM}{"16}
\DeclareMathSymbol{\upartial}{0}{UPM}{"40}
\DeclareMathSymbol{\leqslant}{3}{AMSa}{"36}
\DeclareMathSymbol{\geqslant}{3}{AMSa}{"3E}
\fi
\fi
\fi 

\ifCUPmtlplainloaded \else
\ifAMStwofonts \else 
\def\upi{\pi}
\def\umu{\mu}
\def\upartial{\partial}
\fi
\fi

\title[X--ray cluster properties] {X--ray properties of galaxy
  clusters and groups from a cosmological hydrodynamical simulation}
  \author[S. Borgani et al.]  {S. Borgani$^{1,2}$, G. Murante$^3$,
  V. Springel$^4$, A. Diaferio$^5$, K. Dolag$^6$, \\~\\
\LARGE{L. Moscardini$^7$, G. Tormen$^6$, L. Tornatore$^1$, 
  P. Tozzi$^8$} \\~\\ 
$^1$ Dipartimento di Astronomia dell'Universit\`a di Trieste, via
  Tiepolo 11, I-34131 Trieste, Italy (borgani,tornatore@ts.astro.it)\\
$^2$ INFN -- National Institute for Nuclear Physics, Trieste,
  Italy\\ 
$^3$ INAF, Osservatorio Astronomico di Torino, Strada Osservatorio 20,
  I-10025 Pino Torinese, Italy (giuseppe@to.astro.it)\\
$^4$ Max-Planck-Institut f\"ur Astrophysik, Karl-Schwarzschild Strasse
  1, Garching bei M\"unchen, Germany (volker@mpa-garching.mpg.de)\\
$^5$ Dipartimento di Fisica Generale ``Amedeo Avogadro'', Universit\'a
  degli Studi di Torino, Torino, Italy (diaferio@ph.unito.it) \\
$^6$ Dipartimento di Astronomia, Universit\`a di Padova, vicolo
  dell'Osservatorio 2, I-35122 Padova, Italy (kdolag,tormen@pd.astro.it)\\
$^7$ Dipartimento di Astronomia, Universit\`a di Bologna, via Ranzani
  1, I-40127 Bologna, Italy (moscardini@bo.astro.it)\\
$^8$ INAF, Osservatorio Astronomico di Trieste, via Tiepolo 11,
  I-34131 Trieste, Italy (tozzi@ts.astro.it)}

\pubyear{2003}

\begin{document}
\label{firstpage}
\maketitle

\begin{abstract}
We present results on the X--ray properties of clusters and groups of
galaxies, extracted from a large cosmological hydrodynamical
simulation. We used the Tree+SPH code {\tt GADGET} to simulate a
concordance $\Lambda$CDM cosmological model within a box of $192\hm$
on a side, $480^3$ dark matter particles and as many gas
particles. The simulation includes radiative cooling assuming
zero metallicity, star formation and supernova feedback. The very
high dynamic range of the simulation allows us to cover a fairly large
interval of cluster temperatures.  We compute X--ray observables of
the intra--cluster medium (ICM) for simulated groups and clusters and
analyze their statistical properties. The simulated mass--temperature
relation is consistent with observations once we mimic the procedure
for mass estimates applied to real clusters. Also, with the adopted
choices of $\Omega_m=0.3$ and $\sigma_8=0.8$ for matter density and
power spectrum normalization, respectively, the resulting X--ray
temperature function agrees with the most recent observational
determinations. The luminosity--temperature relation also agrees with
observations for clusters with $T\magcir 2$ keV. At the scale of
groups, $T\mincir 1$ keV, we find no change of slope in this
relation. The entropy in central cluster regions is higher than
predicted by gravitational heating alone, the excess being almost the
same for clusters and groups.  We also find that the simulated
clusters appear to have suffered some overcooling. We find $f_*\simeq
0.2$ for the fraction of baryons in stars within clusters, thus about
twice as large as the value observed.  Interestingly, temperature
profiles of simulated clusters are found to steadily increase toward
cluster centers.  They decrease in the outer regions, much like
observational data do at $r\magcir 0.2\,r_{\rm vir}$, while not
showing an isothermal regime followed by a smooth temperature decline
in the innermost regions.  Our results thus demonstrate the need for
yet more efficient sources of energy feedback and/or the need to
consider additional physical process which may be able to further
suppress the gas density at the scale of poor clusters and groups,
and, at the same time, to regulate the cooling of the ICM in central
regions.

\end{abstract}

\begin{keywords}
Cosmology: numerical simulations -- galaxies: clusters --
hydrodynamics -- $X$--ray: galaxies
\end{keywords}

\section{Introduction}
Observations of galaxy clusters in the X--ray band offer a unique
means to study the physical properties of the diffuse cosmic baryons
in the intra--cluster medium (ICM; see Sarazin 1988, for a historical
review). Under the action of gravity, these baryons follow the dark
matter during the process of hierarchical structure formation, in
which they are heated by adiabatic compression during the halo mass
growth and by shocks induced by supersonic accretion or merger events.
Since gravity does not have any preferred scale, clusters and groups
are in principle expected to appear as scaled version of each other,
provided gravity dominates the process of gas heating (Kaiser 1986)
and the power spectrum of primordial perturbations is featureless over
the relevant scales.  Under the additional assumptions that gas is in
hydrostatic equilibrium within the dark matter (DM) potential wells
and that bremsstrahlung dominates the emissivity, this scenario
predicts self--similar X--ray scaling relations for cluster and group
properties: {\em (i)} $L_X\propto T^2$ for the relation between X--ray
luminosity and gas temperature; and {\em (ii)} $M_{\rm gas}\propto
M_{\rm vir}\propto T^{3/2}$ for the relation between gas mass, total
virialized mass and temperature. Furthermore, if we define the gas
entropy as $S=T/n_e^{2/3}$, then the self--similarity of gas density
profiles leads to the scaling $S\propto T$ if entropy is estimated at
a fixed overdensity for different clusters. The overall validity of
these scaling relations has been confirmed by hydrodynamical
simulations of galaxy clusters that included only gravitational
heating (e.g., Navarro, Frenk \& White 1995; Evrard, Metzler \&
Navarro 1996; Bryan \& Norman 1998; Eke, Navarro \& Frenk 1998).

However, a variety of observational evidences demonstrates that this
simple picture does not apply to real clusters.  The
luminosity--temperature relation is observed to be steeper than
predicted, $L_X\propto T^\alpha$, with $\alpha \simeq 2.5$--3 for
clusters with $T> 2$ keV (e.g., White, Jones \& Forman 1997;
Markevitch 1998; Arnaud \& Evrard 1999; Ettori, De Grandi \& Molendi
2002), with indications of an even steeper slope at the scale of
groups, $T\mincir 1$ keV (e.g., Ponman et al. 1996; Helsdon \& Ponman
2000; Sanderson et al. 2003; cf. Mulchaey \& Zabludoff 1998, and
Osmond \& Ponman 2003). In addition, the evolution of this relation
appears to be slower than predicted by self--similarity (e.g., Holden
et al. 2002; Novicki, Sornig \& Henry 2002; Ettori et al. 2003, and
references therein) although this result is still a matter of debate
(e.g., Vikhlinin et al. 2002). Also, the relation between gas mass and
temperature is observed to be steeper than the self--similar one,
$M_{\rm gas}\propto T^\alpha$, with $\alpha\simeq 1.7$--2.0 (e.g.,
Mohr, Mathiesen \& Evrard 1999; Vikhlinin, Forman \& Jones 1999;
Neumann \& Arnaud 1999; Ettori et al. 2002b), or, equivalently, poor
clusters and groups contain a relatively smaller amount of gas (e.g.,
Sanderson et al. 2003, and references therein). Finally, the gas
entropy within clusters is in excess with respect to what is expected
from self--similar scaling (e.g., Ponman, Cannon \& Navarro 1999;
Lloyd--Davies, Ponman \& Cannon 2000; Finoguenov et al. 2002), with a
dependence on temperature roughly equal to $S\propto T^{2/3}$ (Ponman,
Sanderson \& Finoguenov 2003).

These observational results indicate that non--gravitational processes
that took place during cluster formation must have substantially
affected the physics of the ICM and left an imprint on its X--ray
properties. A variety of models have been developed so far to explain
the resulting ICM properties and, in particular, the lack of
self--similarity between clusters and groups.  These models can be
broadly classified into two categories: those which are based on
non--gravitational heating processes of the ICM (e.g., Evrard \& Henry
1991; Kaiser 1991; Bower 1997; Cavaliere, Menci \& Tozzi 1998; Balogh,
Babul \& Patton 1999; Tozzi \& Norman 2001; Babul et al. 2002), and
those which resort on the effects of radiative cooling (e.g., Bryan
2000; Voit \& Bryan 2001; Wu \& Xue 2002; Voit et al. 2002).

Non--gravitational heating increases the entropy of the gas, which can
prevent it from reaching high density during the cluster collapse. If
a given amount of heating energy per particle, say $E_h$, is assigned
to the gas, then we expect the effect of extra heating to be
negligible for massive clusters with virial temperature $T>E_h$, while
it should leave a significant imprint on smaller systems with
$T\mincir E_h$. As a consequence, X--ray luminosity and gas mass are
suppressed by a larger amount in smaller systems, thus causing a
steepening of the $L_X$--$T$ and $M_{\rm gas}$--$T$
relations. Furthermore, extra heating also sets a minimum value for
the entropy that gas can reach in central regions of clusters and
groups, which can in principle account for the observed excess. In
fact, both semi--analytical models (e.g., Balogh et al. 1999; Tozzi \&
Norman 2001) and numerical simulations (e.g., Bialek et al. 2001;
Brighenti \& Mathews 2001; Borgani et al. 2001a, 2002) were able to
demonstrate that observational data can be reproduced by assigning a
heating energy of about 0.5--1 keV per particle or, equivalently, by
imposing a pre--collapse entropy floor of $S_{\rm fl}\sim 50$--100 keV
cm$^2$. 

As for the origin of extra--heating, supernovae (SN) have been
considered as a first possibility (e.g., Menci \& Cavaliere 2000;
Bower et al. 2001). Using the metal content of the ICM as a diagnostic
for the number of SN exploded (e.g., Renzini 1997, 2003; Kravtsov \&
Yepes 2000; Finoguenov, Arnaud \& David 2001; Pipino et al. 2002),
several authors concluded that SN may however not be able to supply
the required amount of feedback energy. A pristine, essentially
metal-free stellar population, the so--called Pop-III stars, has also
been suggested to contribute significantly to pre--heating
(Loewenstein 2001), although their contribution is constrained by the
requirement of not to over--heat and over--pollute the high--redshift
intergalactic medium (IGM; Scannapieco, Schneider \& Ferrara
2003). Another possible source for ICM heating is represented by AGN
(e.g, Valageas \& Silk 1999; Wu, Fabian \& Nulsen 2000; Mc Namara et
al. 2000; Cavaliere, Lapi \& Menci 2002). In this case, the available
energy budget is in principle quite large, but a coherent treatment of
the conversion of mechanical energy of jets into thermal energy of the
diffuse medium (Reynolds, Heinz \& Begelman 2002; Omma et al. 2003) is
still missing.

Although it may sound like a paradox, radiative cooling has also the
effect of increasing the entropy of the diffuse cluster baryons. This
results as a consequence of the selective removal of low--entropy gas,
which is characterized by a cooling time shorter than the typical
cluster age (e.g., Voit \& Bryan 2001; Wu \& Xue 2002). Besides
increasing the observed mean entropy, the removal of gas
from the X--ray emitting phase also reduces the X--ray luminosity
(e.g., Muanwong et al. 2002; Dav\'e et al. 2002), much like in the
heating scenario. Although cooling must clearly occur at some level as
soon as gas reaches high density within collapsed halos, it has the
unpleasant feature to be a runaway process. This manifests itself in numerical simulations
that include gas cooling and star formation, but no efficient heating
processes. These simulations invariably find that a very large
fraction $f_*$ of gas is converted into a ``stellar'' cold medium,
with $f_*\magcir 30$ per cent (e.g., Suginohara \& Ostriker 1998;
Lewis et al. 2000; Yoshida et al. 2002; Tornatore et al. 2003), which
lies substantially above typical observed values, $f_*\mincir 10$ per
cent (e.g., Balogh et al. 2001; Lin, Mohr \& Stanford 2003), derived
from the local luminosity density of stars.

This demonstrates the need to develop a more realistic and
self--consistent description of the ICM where the effect of cooling is
counteracted and regulated by energy feedback from astrophysical
sources (e.g., Oh \& Benson 2003).  In this spirit, Voit et al. (2002)
have developed a semi--analytical framework which includes the
combined effect of cooling and extra heating. While cooling is
responsible for setting the level of the entropy floor in this model,
extra heating regulates the amount of gas which lies above the entropy
limit for the onset of cooling. However, combining heating and cooling
in a dynamically self--consistent way has been not achieved
yet. Hydrodynamical simulations of clusters including cooling and
different models for non--gravitational heating (Tornatore et
al. 2003) have shown that these two effects interact with each other
in a non--trivial way and, in general, it is not at all obvious that
they can be combined such that overcooling is avoided while
simultaneously providing a good fit to the X--ray scaling relations.

As the level of complexity in the description of the ICM physics is
increased, hydrodynamical simulations are becoming invaluable
theoretical tools to keep pace with the observational progress brought
about by the unprecedented quality of X--ray data from the Chandra and
XMM--Newton satellites. However, an important factor limiting the
reliability of numerical simulations is given by their numerical
resolution, which is usually determined by a combination of the
available supercomputing time and the simulation code's capabilities.
For this reason, numerical studies of the ICM typically represent a
compromise between the mass resolution that one wants to achieve
within each single cluster--sized halo and the number of clusters and
groups that one wants to study numerically. Finding an optimal
compromise is not easy when one is interested in X--ray studies of
clusters. On one hand, the dependence of the bremsstrahlung emissivity
on the density squared requires that small--scale details of the gas
distribution are correctly represented, otherwise the simulated X-ray
emissivity will be incorrect. On the other hand, a reliable comparison
with observational results on cluster X--ray scaling relations
requires a statistically representative ensemble of halos to be
simulated, which can only be obtained in a large simulation volume,
at the price of compromising the mass-resolution.

Hydrodynamical simulations of individual cluster--sized halos, based
on zoom--in resimulation techniques (e.g., Katz \& White 1993; Tormen,
Bouchet \& White 1997), presently allow each object to be represented
with $\sim 10^5$ particles, with a force resolution of about 5
$h^{-1}$kpc (e.g., Borgani et al. 2002; Valdarnini 2003; Tornatore et
al. 2003; Tormen, Moscardini \& Yoshida 2003). On the other hand,
simulations of cosmological boxes, with sizes ranging from about 50 up
to few hundreds $\hm$ on a side, have been run with the purpose of
simulating in one realization a statistically representative number of
clusters and groups (e.g., Bryan \& Norman 1998; Muanwong et al. 2002;
Bond et al. 2002; Dav\'e et al. 2002; Zhang, Pen \& Wang 2002; White,
Hernquist \& Springel 2003; Springel \& Hernquist 2003b; Motl et
al. 2003). For example, Muanwong et al. (2002) analysed X--ray
properties of clusters and groups in simulations with a box--size of
100$\hm$ containing $2\times 160^3$ DM and gas particles, with
gravitational softening of a few tens $h^{-1}$ kpc. In order to achieve a
better mass and force resolution, Dav\'e et al. (2002) and Kay et
al. (2003) adopted a smaller box size of 50$\hm$ with $2\times 144^3$
and $2\times 128^3$ gas and DM particles, respectively, thus
restricting themselves to the study of galaxy groups; rich clusters
are not found in such small volumes. While these simulations were
based on the SPH technique, Motl et al. (2003) used an Eulerian code
capable of adaptive mesh refinement (AMR) to simulate a cosmological
box of 256$\hm$ on a side, reaching a mass resolution of about
$10^{10}h^{-1}M_\odot$ at their highest refinement level. This
simulation included radiative cooling, but neglected the effect of SN
feedback.

In this paper, we present results on the X--ray properties of clusters
and groups identified in a new, very large SPH simulation within a
cosmological box of size 192$\hm$ on a side, using $480^3$ DM
particles and as many gas particles. The simulation includes radiative
cooling, a prescription for star formation in a multi--phase model for
the interstellar medium (ISM), and a recipe for galactic winds
triggered by SN explosions, as described in full detail by Springel \&
Hernquist (2003a). Thanks to the force and mass resolution achieved,
we resolve galaxy groups, having a temperature of about 0.5 keV, with
$\sim 5000$ gas particles, while the most massive halos found in the
box have $\sim 10^5$ gas particles within the virial radius.  This
simulation hence combines fairly high resolution in a large
cosmological volume with a quite advanced treatment of the gas
physics. It is thus ideally suited for a comparison between simulated
and observed X--ray properties of groups and clusters, allowing us to
shed more light on the interplay between the properties of the ICM and
the processes of star formation in cluster galaxies.

The outline of the paper is as follows. In Section~2, we describe the
numerical method that we use, and provide an overview of the general
characteristics of the simulation. Section~3 is devoted to the
presentation of the results. After discussing the star fraction
produced within the cluster regions, we study the different X--ray
observables, such as luminosity, temperature and entropy. Much
emphasis will be given through all of this section to a comparison of
the numerical results with X--ray observations. Finally, we summarize
our results and draw our main conclusions in Section~4.

\section{The simulation}

The cosmological model we simulated represents a standard flat
$\Lambda$CDM universe, with matter density $\Omega_m=0.3$, Hubble
constant $H_0=70$ km s$^{-1}$Mpc$^{-1}$, baryon density $\Omega_{\rm
bar}=0.04$ and normalization of the power spectrum
$\sigma_8=0.8$. This normalization is somewhat lower, though
consistent within 1$\sigma$, than suggested by the WMAP result
(Spergel et al. 2003), but it is consistent with recent determinations
based on the number density of galaxy clusters (e.g., Pierpaoli et
al. 2003, and references therein), or based on cosmic shear
measurements (see Refregier 2003, for a review). The baryon density
agrees with the prediction of big--bang nucleosynthesis for the
deuterium abundance found in high--$z$ Lyman--$\alpha$ clouds by
Kirkman et al. (2003), while it is $\simeq 20$ per cent lower, though
consistent at about $1\sigma$ level, with the WMAP value.

Initial conditions have been generated at redshift $z_{\rm
start}\simeq 46$ within a box of $192\hm$ on a side using the {\tt
COSMICS} package provided by
E. Bertschinger\footnote{http://arcturus.mit.edu/cosmics/}. The
density field has been sampled with $480^3$ dark matters and an equal
number of gas particles, with masses of $m_{\rm DM}=6.6\times
10^9M_\odot$ and $m_{\rm gas}=9.9\times 10^8M_\odot$, respectively.
For our choices of starting redshift and mass resolution, the rms
Zeldovich displacement in the initial conditions was equal to about
one tenth of the mean interparticle separation. During the evolution,
the number of gas particles changes as a consequence of their partial
conversion into new star particles. Because the generated star
particles have mass smaller than gas particles, the total number
of particles actually increases by a small amount as a result of star
formation.

The run has been realized using {\small GADGET}\footnote{\tt
http://www.mpa-garching.mpg.de/gadget}, a massively parallel tree
N--body/SPH code (Springel, Yoshida \& White 2001) with fully adaptive
time--step integration. We here used {\small GADGET-2}, a new version
of this simulation code that is more efficient than earlier versions
of {\small GADGET}, and offers better time--stepping for collisionless
dynamics, among other improvements.  Of particular relevance is the
implementation of SPH adopted in the code, which follows the
formulation suggested by Springel \& Hernquist (2002). This method
explicitly conserves energy {\em and} entropy, where appropriate, and
substantially reduces numerical overcooling problems at interfaces
between hot and cold gas. Optionally, the new code also allows the use
of a TreePM algorithm (Bagla 2002) to speed up the computation of the
long-range gravitational force, an approach that we employed in the
present simulation.

Radiative cooling was computed assuming an optically thin gas of
primordial composition (mass--fractions of $X=0.76$ for hydrogen and
$1-X=0.24$ for helium) in collisional ionization equilibrium,
following Katz, Weinberg \& Hernquist (1996).  We have also included a
photoionizing, time--dependent, uniform UV background expected from a
population of quasars (e.g., Haardt \& Madau 1999), which reionizes
the Universe at $z\simeq 6$.  The effect of a photoionizing background
is that of inhibiting gas collapse and subsequent star formation
within the halos of sub--$L_*$ galaxies (e.g., Benson et al. 2002),
thus having a secondary impact at the resolution of our simulation.
Although the code includes a method to follow metal production (see
below), we have not included the effects of metals on the cooling
function, owing both to code limitations and to the approximate
treatment of metal generation and diffusion.

\begin{figure*}
\centerline{
\hbox{
\psfig{file=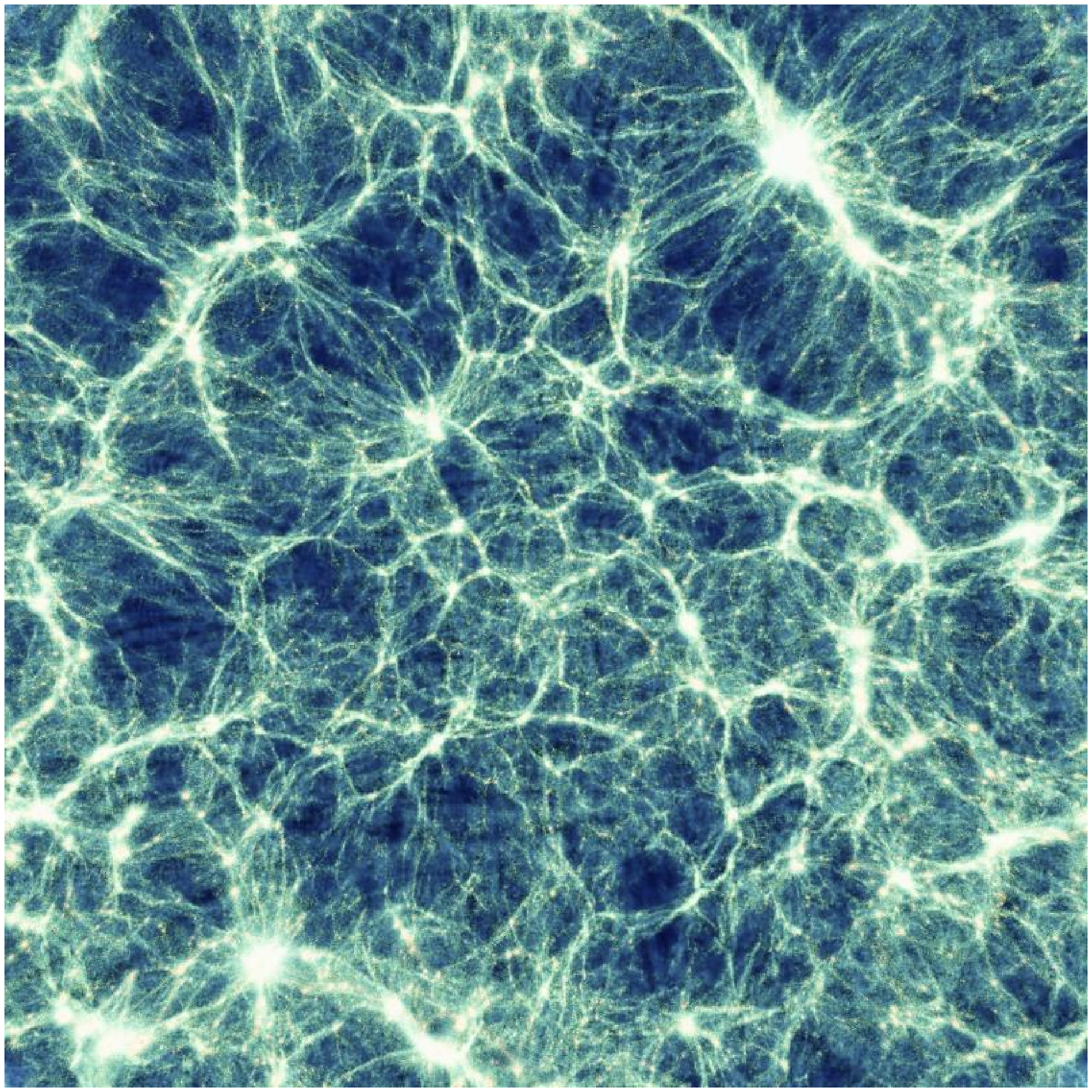,width=8.cm} 
\psfig{file=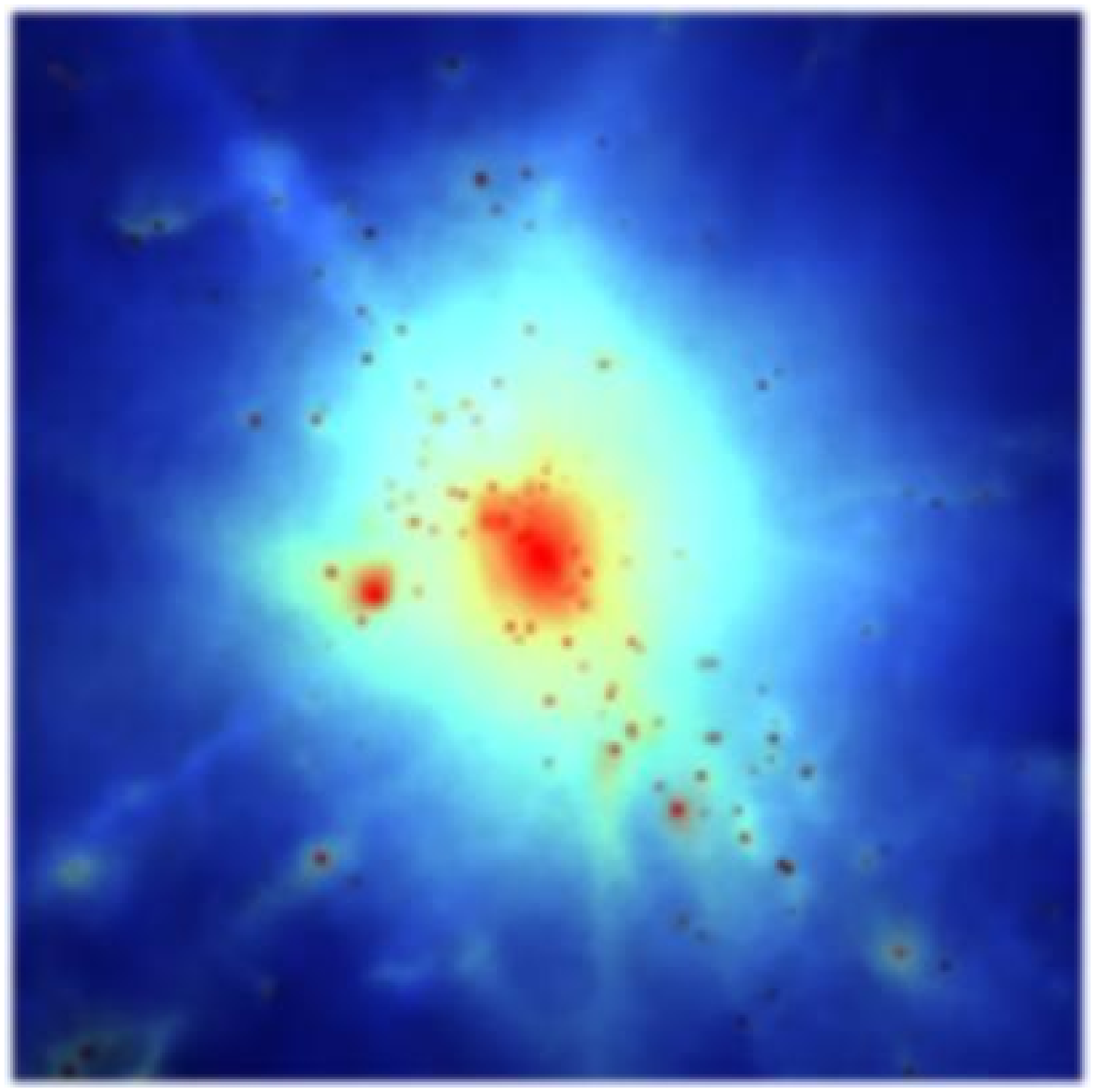,width=8.15cm} 
}
}
\caption{Left panel: map of the gas density over the whole simulation
  box at $z=0$, projected using a ray--tracing technique through a
  slice having thickness of 12 $\hm$, and containing the most massive
  cluster found in the simulation (upper right side of the
  panel). Right panel: zoom into the region of the largest cluster;
  the cluster is shown out to one virial radius, so that the panel
  encompasses a physical scale of about 4.5 $h^{-1}$ Mpc.}
\label{fi:maps} 
\end{figure*}

Star formation is treated using the hybrid multiphase model for the
interstellar medium introduced by Springel \& Hernquist (2003a).  We
refer to this paper for a detailed description of the method,
providing here only a short summary of the model. The ISM is pictured
as a two-phase fluid consisting of cold clouds that are embedded at
pressure equilibrium in an ambient hot medium.  The clouds form from
the cooling of high density gas, and represent the reservoir of
baryons available for star formation. When stars form, according to a
Salpeter IMF (Salpeter 1955), the energy released by supernovae heats
the ambient hot phase of the ISM, and in addition, clouds in supernova
remnants are evaporated. These effects establish a tightly
self--regulated regime for star formation in the ISM.  In practice,
the scheme is numerically implemented as a sub-resolution model,
i.e.~cold clouds are not resolved individually. Instead, only their
total mass fraction in each element of the ISM is computed, otherwise
they are treated in a stochastic fashion with their collective effect
on the ISM dynamics being described by an effective equation of state.
The numerical implementation of this multiphase model describes each
gas particle as composed by a hot component, having its own mass and
density, and a cold neutral component. The relative amount of these
two phases is determined by the local value of gas density and
temperature.  We note that metal enrichment and type-II supernovae
feedback is computed assuming an instantaneous recycling
approximation.

If not counteracted by some sort of feedback process, cooling is well
known to overproduce the amount of stars both in the average
environment and in the group/cluster overdense environment
(e.g. Balogh et al. 2001, and references therein).  As discussed by
Springel \& Hernquist (2003a), their multiphase ISM model alone does
however not fully resolve this problem, despite its ability to
regulate the consumption of cold gas into stars within the ISM. This
is because the cooling rates within halos remain essentially
unaffected in the model, i.e.~the supply of gas to the dense
star-forming ISM is largely unchanged, while by construction the
phases of the ISM remain coupled to each other, preventing baryons to
leave the ISM (except for dynamical effects like gas stripping in
mergers).

However, galactic outflows are observed and expected to play a key
role in transporting energy and metals produced by the stellar
population into the IGM/ICM. To account for them, Springel \&
Hernquist (2003a) suggested a phenomenological description of galactic
winds as an extension of their model, and we have included such winds
in our simulation.  According to our choice of parameters for the
feedback and wind scheme, star-forming gas particles contribute to the
wind with a mass outflow rate two times larger than their star
formation rate, with a wind velocity of about $360\,{\rm km\,
s^{-1}}$. This velocity is less extreme than the value of about
$480\,{\rm km\, s^{-1}}$ adopted by Springel \& Hernquist (2003b) in
their comprehensive study of the cosmic star formation history.  In
fact, the energy of our winds is only half that of Springel \&
Hernquist (2003b), who assumed that essentially all of the feedback
energy is available to power the winds. With this choice, Springel \&
Hernquist (2003b) were able to show that the global efficiency of
cooling and star formation is reduced to the observed level, and that
observational constraints on the amount of neutral hydrogen in high
column--density absorbing systems at high redshift (Nagamine, Springel
\& Hernquist 2003a,b) can be accounted for. However, we preferred here
a more conservative choice of less energetic winds, so as to allow for
radiative losses taking place in the interstellar medium at
sub--resolution scales.

The Plummer--equivalent gravitational softening of the simulation was
set to $\epsilon_{\rm Pl}=7.5\,h^{-1}$ kpc comoving from $z=2$ to
$z=0$, while it was taken to be fixed in physical units at higher
redshift. We used 32 neighboring particles for the SPH computations,
but did not allow the SPH smoothing length to drop to less than one
quarter of the value of the gravitational softening length of the gas
particles. In total, the simulation required about 40,000 CPU hours on
64 processors of the IBM-SP4 machine located at CINECA. We produced
100 snapshots at log-equispaced values of the expansion factor, from
$a_{\rm exp}=0.1$ to $a_{\rm exp}=1$, thus producing a total amount of about
1.2 Tb of data. The fine spacing of snapshots in time can be used to
measure merger tree of halos, and to realize projections along the
backward light--cone.

\section{Results}

We start our analysis with an identification of groups and clusters
within the simulations box.  To this end, we first apply a
friends-of-friends halo finder to the distribution of DM particles,
with a linking length equal to 0.15 times their mean separation. For
each group of linked particles with more than 500 members, we identify
the particle having the minimum value of the gravitational
potential. This particle is then used as a starting point to run a
spherical overdensity algorithm, which determines the radius around
the target particle that encompasses an average density equal to the
virial density for the adopted cosmological model, $\rho_{\rm
vir}(z)=\Delta_c(z)\rho_{c}(z)$, where
$\rho_c(z)=[H(z)/H_0]^2\rho_{c,0}$ is the critical density at redshift
$z$, and the overdensity $\Delta_c(z)$ is computed as described in
Eke, Cole \& Frenk (1996).

In the left panel of Figure \ref{fi:maps}, we show a map of the gas
distribution at $z=0$, projected through a slice of thickness 1/16th
of the box size. This slice includes the most massive cluster found in
the simulation, which is located in the upper right region of the
map. The panel on the right shows the gas distribution of this cluster
out to the virial radius, thus representing a zoom-in by about a factor
of 40. The tiny dark spots visible in the central cluster regions
are condensations of high--density cold gas. They mark the locations
where star formation is taking place and, therefore, the positions of
cluster galaxies. The amount of small--scale detail which is visible
in the zoom--in of the right panel demonstrates the large dynamic
range encompassed by the hydrodynamic treatment of the gas in our
simulation.

The cluster shown in Fig.~\ref{fi:maps}, which has an
emission--weighted temperature of $T_{\rm ew}\simeq 7\,{\rm keV}$ (see
below), is resolved with about $2\times 10^5$ DM particles within the
virial radius. Overall, we have 400 halos resolved with at least
10,000 DM particles, 72 clusters with $T_{\rm ew}>2$ keV, out of which
23 have $T_{\rm ew}>3$ keV. Clusters with $T_{\rm ew}=1$ keV are
resolved with about 7,000 DM particles. Therefore, our simulation
provides us with an unprecedented large sample of simulated groups and
clusters of medium-to-low richness that are represented with good
enough numerical resolution to obtain reliable estimates of X--ray
observable quantities, such as luminosity, temperature and entropy.

In the following, we will mainly concentrate on the description of the
properties of clusters at $z=0$, while we will deserve to a
forthcoming paper the discussion of the redshift evolution of the
X--ray scaling relations and their comparison with observational data.

\subsection{The stellar fraction in clusters}

\begin{figure}
\centerline{
\psfig{file=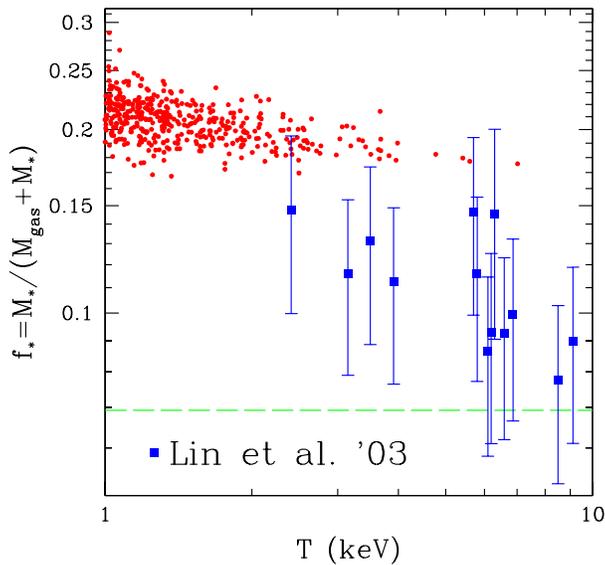,width=8.5cm} 
}
\vspace{-.5truecm}
\caption{The fraction of gas locked into stars, estimated at the
  virial radius from simulations (small circles) and from
  observational data (big squares with errorbars; from Lin et
  al. 2003). The horizontal dashed line indicates the cosmic value of
  $f_*$ found in the simulation.
}
\label{fi:fstar} 
\end{figure}

Observational determinations of the fraction of baryons locked up in
stars in galaxy clusters, $f_*$, consistently indicate a rather small
value. For instance, Balogh et al. (2001) found this fraction to be
below 10 per cent, independent of the cluster richness. More recently,
Lin et al. (2003) selected a sample of nearby clusters, with ICM
masses available from ROSAT-PSPC data and total stellar mass estimated
from the total K--band luminosity, as provided by the 2MASS survey
(e.g., Cole et al. 2001). They found that rich clusters have
$f_*\mincir 10$ per cent, with an increasing trend toward poorer
systems (see the data points plotted in Figure \ref{fi:fstar}).

On the other hand, hydrodynamical simulations of clusters that include
radiative cooling and star formation, have demonstrated that
significantly higher $f_*$ values, as high as 30--50 per cent, are
found when sufficiently high numerical resolution is adopted (e.g.,
Balogh et al. 2001; Dav\'e et al. 2002; Tornatore et al. 2003).  This
reflects the well established result that radiative cooling converts
a large fraction of baryons into collisionless stars (e.g., Suginohara
\& Ostriker 1998; Balogh et al. 2001) if it is not efficiently
counteracted by some sort of feedback process that heats the gas
surrounding star--forming regions, thus increasing its cooling time
and preventing a ``runaway'' overcooling.

While consistency with the observed $f_*$ values can eventually be
reached by adopting suitable schemes for gas pre--heating and feedback
(e.g., Muanwong et al. 2002; Kay et al. 2003; Tornatore et al. 2003;
Marri \& White 2003), it is not trivial at all to implement such
schemes in a numerically self--consistent and physically well
motivated way within cosmological simulations of structure formation.
Note that feedback energy associated with star formation is most
plausibly released in star--forming regions, where gas is at high
density and, therefore, has short cooling time. Increasing the thermal
energy of this gas is hence quite ineffective in preventing it from
cooling. For this reason, a physically motivated model for either
preventing such particles from radiating away the heating energy or
for assigning this energy to particles at lower density (which have
longer cooling times) is needed.

The feedback model by Springel \& Hernquist (2003a) included in our
simulation represents a simple physical model of the expected
multi-phase structure of the ISM. This model has well-controlled
numerical properties, in the sense that the star formation rate in a
given halo converges even at moderate numerical resolution, which is
desirable for simulations of hierarchical structure formation. The
modelization of galactic winds accounts for the role of galactic
ejecta, and is crucial for transferring SN energy out of the
star--forming environment itself.  If strong winds are adopted, this
feedback scheme is in fact able to reproduce the observed cosmic
star--formation history and to reduce the total amount of baryons to
about 10 per cent.

However, in our simulation a somewhat weaker wind model was adopted
(see the discussion in Section 2).  With this choice, the total mean
cosmic value of $f_*\simeq 7$ per cent of stars produced in the
simulation is consistent with observational results (e.g., Fukugita,
Hogan \& Peebles 1998; Balogh et al. 2001), although such a low value
may be due to the lack of objects below our resolution limit. At the
same time, the efficiency of star formation within the high--density
environment of clusters is still in excess with respect to
observations. Values of $f_*$ for individual clusters are at the level
of $\simeq 20$ per cent for $T>3$ keV clusters, with a slight tendency
to increase for colder systems, as shown in Figure
\ref{fi:fstar}. 

As a word of caution, it is important to be aware of the resolution
limitations of our simulation. The necessity of keeping the box--size
large enough for our study compromises its mass resolution, even for
the fairly large number of particles followed in this run.  This is
particularly important in the context of feedback by galactic
winds. Whether or not ejecta can escape the gravitational potential of
a galaxy depends primarily on its virial mass: while it is relatively
easy for a wind to escape from small objects, winds are expected to
provide only inefficient feedback in massive galaxies and halos, where
they are confined gravitationally.  Since we can hardly treat the
effect of winds on unresolved, or poorly resolved, small galaxies, we
may be underestimating the effect of winds on our smallest systems.
This is particularly true at high redshift, when large numbers of
small halos start collapsing. In addition, we miss the cumulative
effect of winds during the hierarchical built-up of structure up to
the point where we start resolving massive enough objects. This could
result in an underestimate of the degree of metal dispersal estimated
from the simulation, for example. While we are not carrying out
detailed resolution studies ourselves, these are ultimately needed to
fully eliminate these uncertainties. However, we note that the star
formation rate of well resolved objects is expected have converged in
our simulation, as demonstrated by Springel \& Hernquist (2003b) for
the same feedback model and simulation code used here.

\subsection{Computing X--ray luminosity and temperature}

Since the simulation adopts a zero--metallicity cooling function, we
accordingly compute X--ray emissivity under this
assumption. Therefore, the resulting X--ray luminosity of each cluster
in a given energy band is defined as
\be
L_X\,=\,(\mu m_p)^{-2}\sum_i^{N_{\rm gas}} m_{h,i}\rho_{h,i}
\Lambda(T_i),
\label{eq:lx}
\ee 
where $\Lambda(T)$ is the cooling function in that energy band. In
this equation the sum runs over all $N_{\rm gas}$ gas particles
falling within $r_{\rm vir}$, and $\mu$ is the mean molecular weight
($=0.6$ for a gas of primordial composition), $m_p$ is the proton
mass, $m_{h,i}$ and $\rho_{h,i}$ are the mass and the density
associated with the hot phase of the $i$--th gas particle,
respectively. A distinction between hot and cold phases is only made
for dense star-forming particles, where the adopted multiphase model
allows a computation of the relative contribution of each of these two
phases, which depends on the local density and temperature (Springel
\& Hernquist 2003a).  By its nature, the neutral cold component is
assumed to not emit any X--rays.  In addition, we also exclude those
particles from the computation of X--ray emissivity whose ionized
component has temperature below $3\times 10^4$ K and density $>500
\rho_c(z)$. Particles inside clusters at such low temperature are
usually at very high density (often they are particles just assigned
to the wind by the phenomenological wind model), much higher than the
threshold chosen above. As such, they would provide a significant, but
spurious, contribution to the X--ray luminosity in central cluster
regions if they were X--ray emitting. These particles
occupy a region in the $\rho$--$T$ plane where in principle only gas
should lie that has already cooled, and so this gas should not be
included in the computation of the X--ray emission (Croft et al. 2001;
Kay et al. 2002).

For the cooling function, we assumed the one provided by Sutherland \&
Dopita (1993), computed for zero metallicity.  Then, corrections with
respect to this emissivity and that within finite energy bands were
computed using the {\tt mekal} model built in XSPEC, again assuming
zero metallicity, thus consistent with the cooling function used in
the simulation code. We remind that the simulation also keeps track of
the expected metal production from the formed stars. Since only
type-II SN are included in this treatment, the resulting metallicity
is mainly contributed by oxygen (e.g., Matteucci 2001, and references
therein). On the other hand, measurements of ICM metallicity mostly
refer to iron, because they typically detect the Fe K-shell and
L--shell lines for hot and cold systems, respectively. This is one of
the reasons why we here do not investigate in detail the effect of the
ICM metallicity on cooling and X--ray emissivity. In any case, metal
lines are expected to provide a significant contribution to the
emissivity only at relatively low temperatures, $T\mincir 2$ keV.

\begin{figure}
\centerline{
\psfig{file=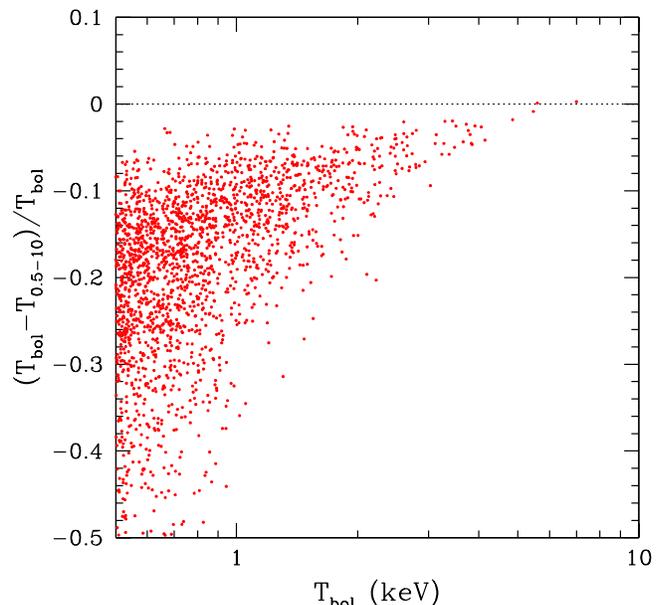,width=8.5cm} 
}
\vspace{-.2truecm}
\caption{The fractional variation of the emission--weighted
  temperature, $T_{\rm ew}$, when using the bolometric and the [0.5--10]
  keV band emissivity.}
\label{fi:tband}
\end{figure}

We define the emission--weighted temperature, $T_{\rm ew}$, as
\be T_{\rm ew}\,=\,{\sum_i^{N_{\rm gas}} m_{h,i}\rho_{h,i}
\Lambda(T_i)\,T_i\over \sum_i^{N_{\rm gas}} m_{h,i}\rho_{h,i} \Lambda(T_i)}\,,
\label{eq:tew}
\ee
and compute it by weighting with the emissivity in the $0.5-10\,{\rm
keV}$ energy band, rather than using bolometric emissivity. This is
meant to reproduce the observational procedure in the estimate of the
temperature from the measured photon spectrum, whose reconstruction at
low energies, say below 0.5 keV, is made hard by instrumental
limitations. As long as the emission--weighted temperature is a good
approximation to the temperature provided by spectral fitting, this
procedure amounts to under-weight low--temperature gas particles.
Therefore, the relevant observable quantity, $T_{0.5-10}$, turns out
to be larger than that based on bolometric emissivity. As shown
in Figure \ref{fi:tband}, this effect is expected to be quite small
for the hottest systems found in the simulation, but can be non
negligible for cold groups, where the temperature estimate is biased
on average to high values by $\sim 20$ per cent.

While all this is true under the assumption of vanishing metallicity,
including metals has the effect of biasing temperatures towards lower
values (Mathiesen \& Evrard 2001). The presence of metals increases
the emissivity of relatively cold particles, thanks to the
contribution of soft lines. Mathiesen \& Evrard (2001) estimated this
effect to lead to an underestimate of the temperature by about 20 per
cent. Since our cooling function has been computed for zero
metallicity, we do not to include this effect here. However, this
emphasizes that a careful like-with-like comparison with observations
is not always straightforward, and requires a careful treatment of
physical processes when running simulations (in this case, the
contribution of metal cooling), and of observational biases when
analyzing them. The contribution from metal lines is expected to affect
estimates of $T_{\rm ew}$ and, to a larger extent, of $L_X$ for
systems with $T\mincir 2$ keV. However, these corrections become
negligible at higher temperatures, where bremsstrahlung dominates the
emissivity.

\subsection{Luminosity profiles}

X-ray surface brightness profiles represent a direct test bed to
establish the existence (or lack) of self--similarity between clusters
and groups. Ponman et al. (1999) actually presented a case against
self--similarity: based on ROSAT-PSPC data, they pointed out that a
continuous change of the mean surface--brightness profiles exists when
going from poor groups to rich clusters. While it was found that this
picture does not necessarily apply to hot ($T\magcir 3$ keV) systems
(e.g., Neumann \& Arnaud 1999), subsequent studies have indeed
confirmed that groups in general show a shallower profile than rich
clusters. This effect is usually quantified by fitting the observed
profiles with standard $\beta$--models for the gas density (Cavaliere
\& Fusco--Femiano 1976),
\be
\rho_{\rm gas}(r)\,=\,{\rho_0\over
  \left[1+(r/r_c)^2\right]^{3\beta_{\rm fit}/2}}\,.
\label{eq:betam}
\ee 
Observationally, $\beta_{\rm fit}$ appears to be an increasing
function of temperature (e.g., Helsdon \& Ponman 2000; Finoguenov et
al. 2001, F01 hereafter; Sanderson et al. 2003).

\begin{figure*}
\centerline{\vbox{
\hbox{
\psfig{file=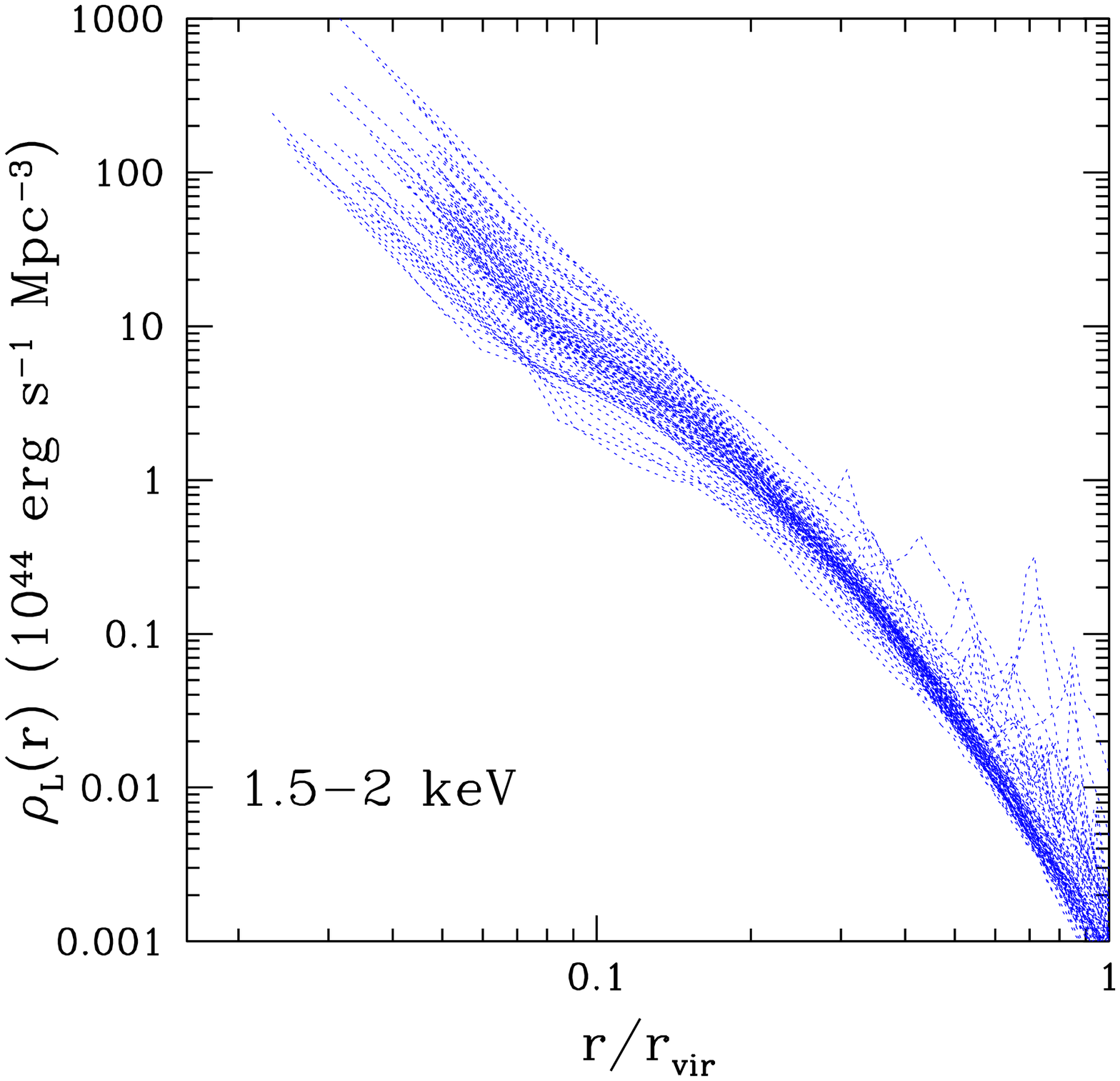,width=6.5cm} 
\psfig{file=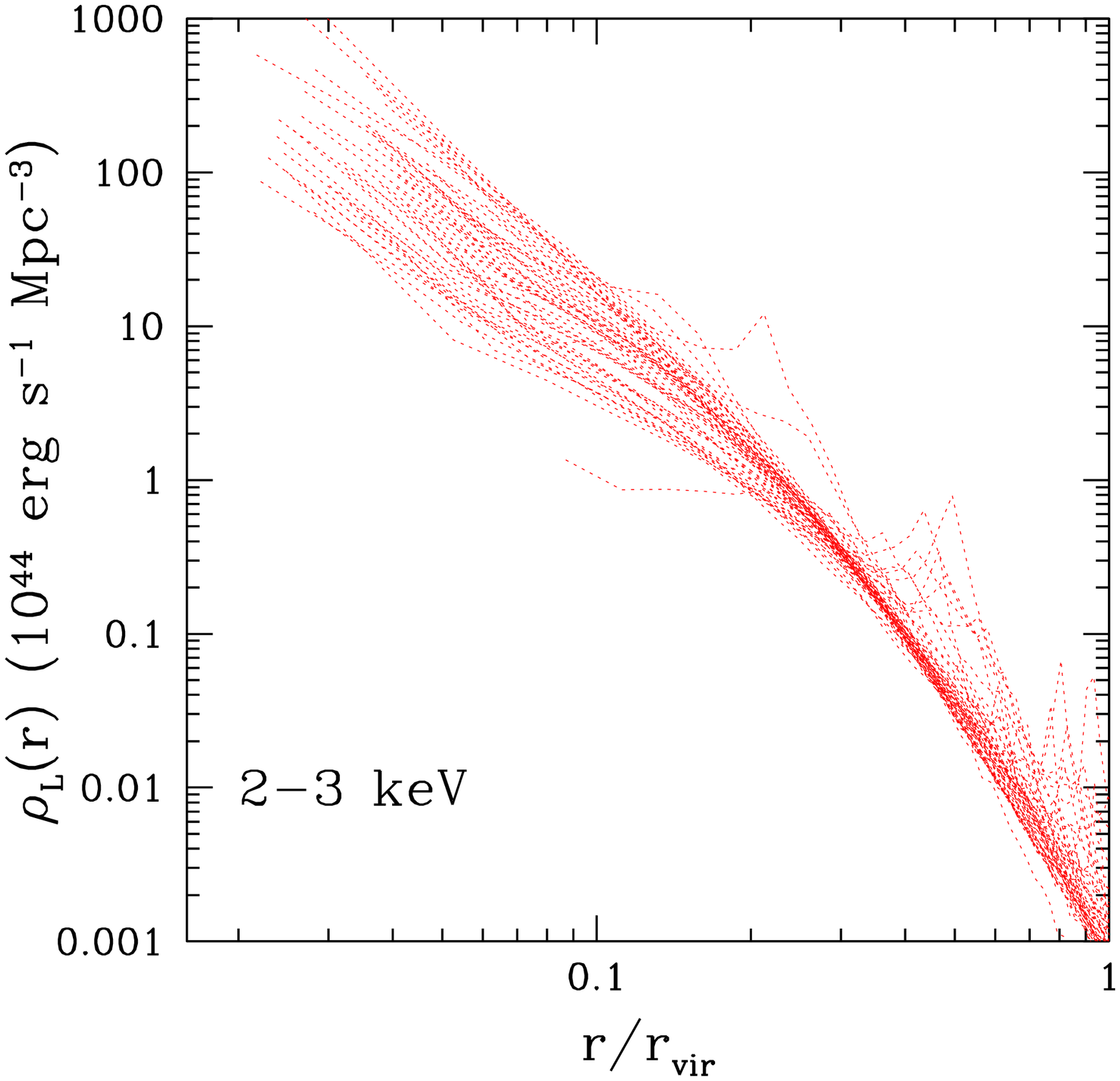,width=6.5cm} 
}
\hbox{
\psfig{file=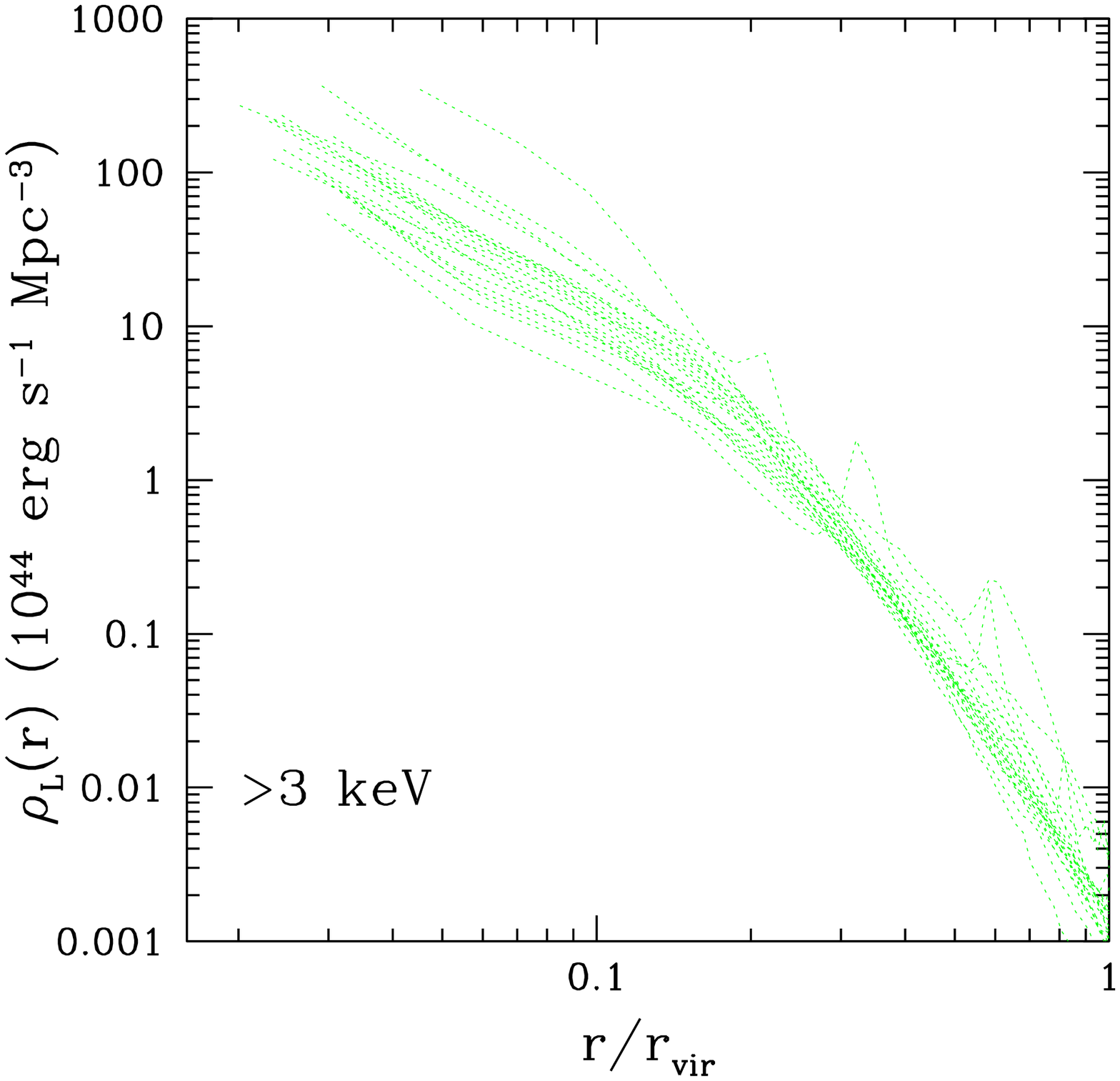,width=6.5cm} 
\psfig{file=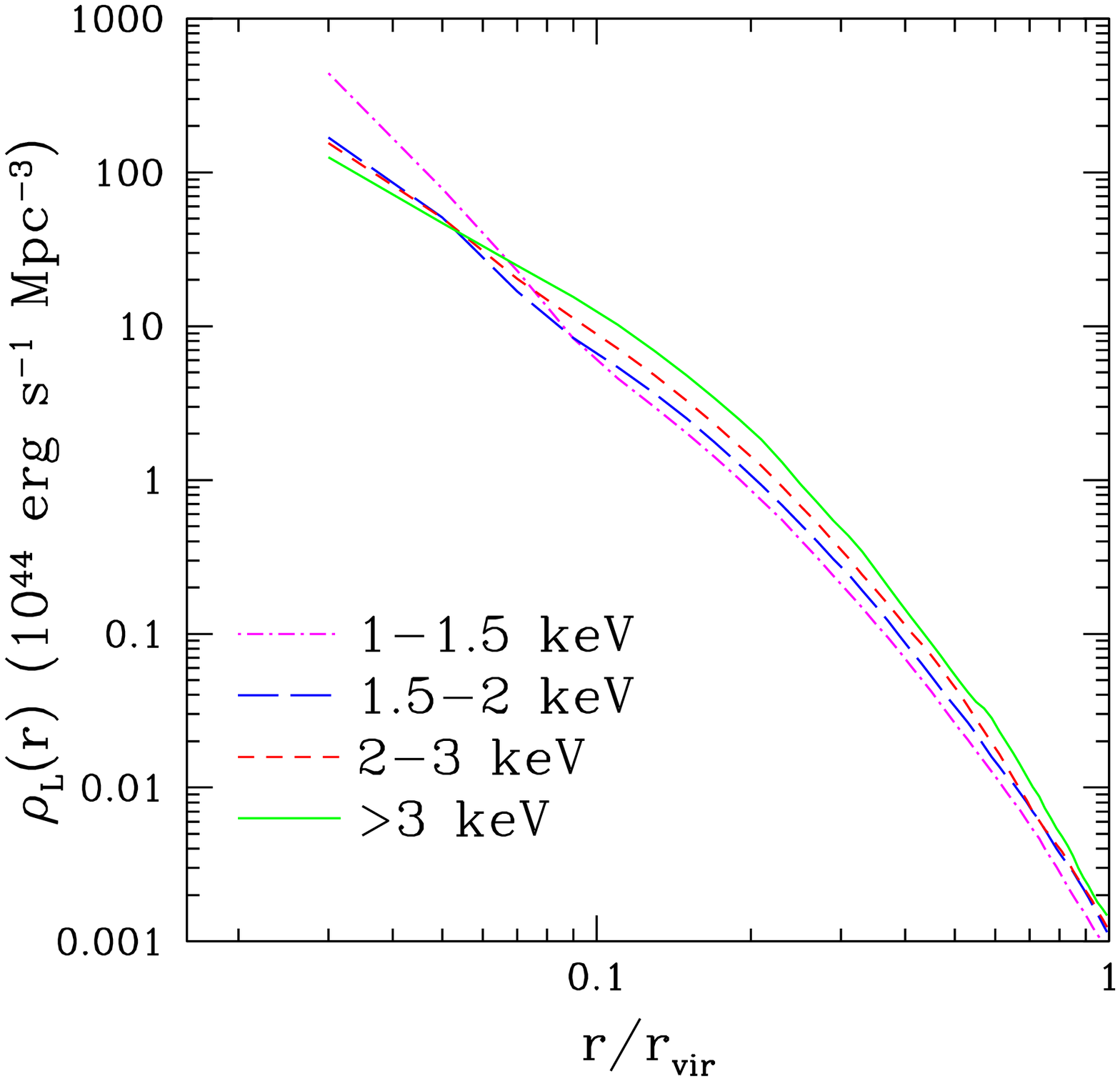,width=6.5cm} 
}
}}
\caption{Profiles of luminosity density at different intervals of
  $T_{\rm ew}$. The upper and the lower left panels show all the profiles
  for structures within different temperature intervals. The lower
  right panel shows the average profiles computed within each range of
  $T_{\rm ew}$.  }
\label{fi:lprof} 
\end{figure*}

In Figure \ref{fi:lprof}, we show the profiles of luminosity density,
$\rho_L(r)$, for groups and clusters within different temperature
intervals. We only compute profiles down to a radius which contains
100 SPH particles. This scale has been shown to be the smallest one
where numerically converged results for the X-ray luminosity can be
computed (Borgani et al. 2002) and, on average, it is about two times
larger than $\epsilon_{\rm soft}=2.8\times \epsilon_{\rm Pl}$, i.e.,
the softening scale where gravitational force starts deviating from
the $1/r^2$ law.

Quite remarkably, once the luminosity profiles are rescaled to the
virial radius, $r_{\rm vir}$, they look very similar in the outer
regions, $r\magcir 0.3r_{\rm vir}$, while the scatter significantly
increases in the innermost regions. This result is in qualitative
agreement with the findings by Neumann \& Arnaud (1999). When looking
at the average profiles (bottom right panel), it is seen that they all
match when the virial radius is approached. However, down to about
$0.1r_{\rm vir}$, colder systems tend to have a slightly shallower
profile, which becomes steeper than that of hot systems at smaller
radii.

The steepening of the gas density profile in the central regions of
groups is the result of the cooling and star--formation process. Since
cooling is relatively more efficient in smaller systems, as also
witnessed by the increasing stellar fraction at low--$T$ (see
Fig. \ref{fi:fstar}), these systems tend to have stronger
compressional heating of gas flowing toward central regions, as a
consequence of the lack of pressure support.  On the other hand,
simulated clusters with $T> 3$ keV do not exhibit any spike in
emissivity associated with their central cooling regions.

\begin{figure}
\centerline{
\psfig{file=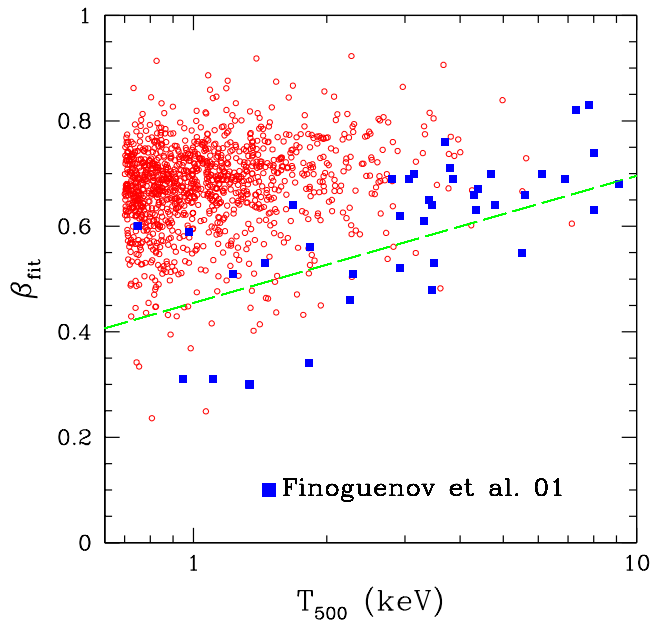,width=8.5cm} 
}
\caption{Comparison between the simulated (open circles) and observed
  (filled squares, from Finoguenov et al. 2001) values of the
  $\beta_{\rm fit}$--parameter obtained by fitting the gas density
  profiles to the $\beta$--model of Eq.~(\ref{eq:betam}). The dashed
  line is the best--fitting model to the data by Sanderson et
  al. (2003), $\beta_{\rm fit}=0.439\,T^{0.20}$.}
\label{fi:bfit} 
\end{figure}

In order to compare our profiles to observational data, we fitted our
simulated gas--density profiles to the $\beta$--model of
Eq.~(\ref{eq:betam}), and compared the result to the compilation of
$\beta_{\rm fit}$ values by F01. Gas density profiles in simulated
clusters are known to progressively steepen towards outer regions,
thus implying that the resulting $\beta_{\rm fit}$ may depend on the
range of scales where the fit is carried out. For this reason, the
range of scales used for the fit should be chosen similar to the one
that was used in the analysis of observational data. As for the outer
radius, we choose to stop at $r_{500}$, which is close to the average
outermost radius used in the analysis of F01, while we adopted
$0.2\,r_{\rm vir}$ for the inner radius, which roughly corresponds to
the radius at which the luminosity profiles change their slopes (see
Fig.~\ref{fi:lprof}).

The result of this analysis is shown in Figure \ref{fi:bfit}, where
the $\beta_{\rm fit}$ values for simulated clusters are compared to
the observational data points by F01 and to the best--fit relation by
Sanderson et al. (2003). The simulation results are consistent with
$\beta_{\rm fit}\simeq 0.7\pm 0.1$, with only a slight tendency for
hotter systems to have larger values of $\beta_{\rm fit}$, a much
weaker trend than inferred from observations. While the resulting
$\beta_{\rm fit}$ values are consistent with the profiles of hot
clusters, simulated groups tend to have steeper gas density profiles
than observed. We verified that this result is robust against
reasonable variations of the range of scales where the fit is
realized.  Reducing the inner radius to the smallest resolved
scale does not produce any appreciable improvement in the comparison
with data, neither it introduces any significant trend for a smaller
$\beta_{\rm fit}$ at the scale of groups. On the other hand,
increasing the outer radius to $r_{200}$ increases the resulting
$\beta_{\rm fit}$ (as first pointed out by Navarro et al. 1995) and,
therefore, makes the disagreement worse. This confirms that neither
cooling nor our description of SN feedback are efficient enough to
reduce the gas density at the center of poor clusters and groups to
the observed level.

A still more faithful comparison with data would require treating
simulated clusters exactly on the same footing as the real ones, by
carefully choosing the same range of scales for sampling the
luminosity profiles.  Vikhlinin et al. (1999) pointed out that fitting
profiles out to $\simeq r_{\rm vir}$ provides slightly larger values
of $\beta_{\rm fit}$, by about 0.05, a trend that is also seen in
simulations. However, such a small increase is not enough for
interpreting the trend of $\beta_{\rm fit}$ with cluster temperature,
as a result of poorer objects being sampled over smaller portions of
their virial regions.  Sanderson et al. (2003) have recently addressed
this point by analysing two clusters for which ROSAT--PSPC data
extends out to quite large radii. They found no significant evidence
of steepening as the fitting region is enlarged, although this
conclusion is based on only two fairly hot systems.  The situation is
likely to improve in coming years as more Chandra and XMM--Newton data
will accumulate, allowing better sampling of the gas density in the
outer regions of clusters and groups.

\subsection{Temperature profiles}
\begin{figure*}
\centerline{\hbox{
\psfig{file=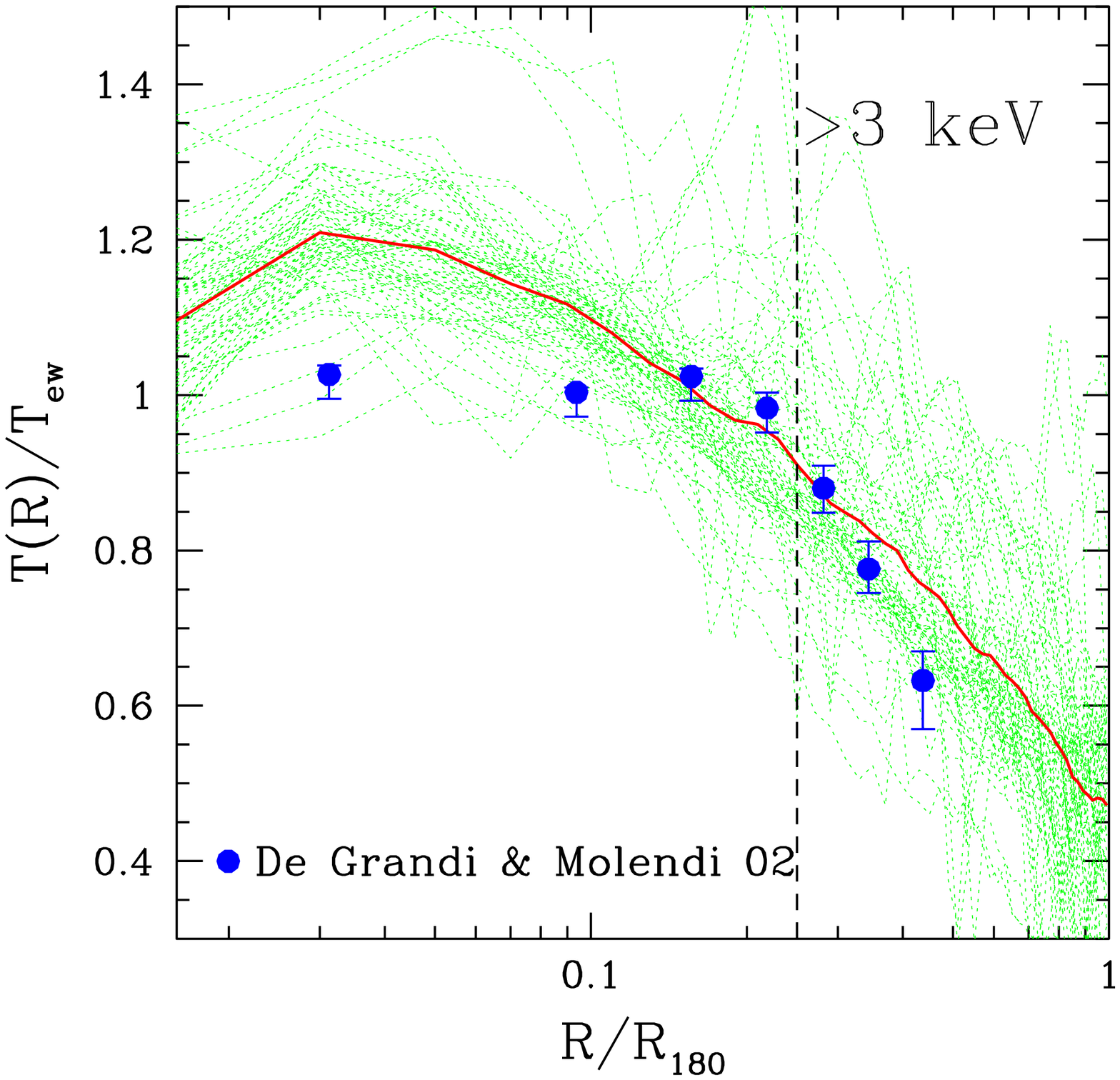,width=7.cm} 
\psfig{file=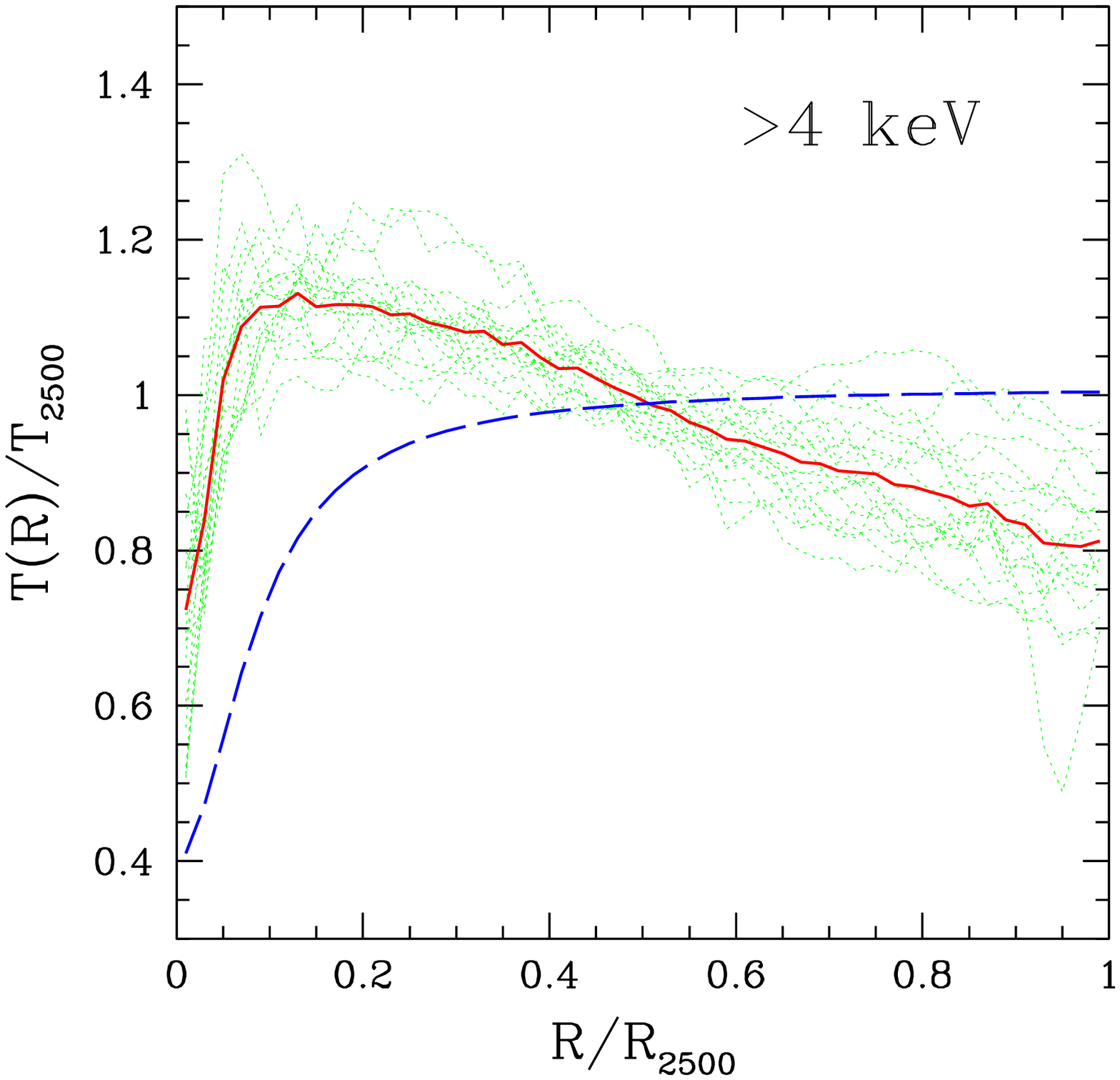,width=7.cm} 
}}
\caption{Comparison between simulated and observed projected
  temperature profiles. Left panel: comparison between simulated
  clusters with $T_{\rm ew}> 3$ keV and the observational data points
  from the analysis of Beppo--SAX data for 17 clusters by De Grandi \&
  Molendi (2002); projected radial scales are in units of $R_{180}$,
  i.e. the radius at which $\bar\rho/\rho_{\rm cr}=180$. Right panel:
  comparison between simulated clusters with $T_{\rm ew}> 4$ keV and
  the best--fitting universal temperature profiles measured by Allen
  et al. (2001) from their analysis of Chandra data for a set of six
  relaxed clusters (dashed curve); projected radial scales are in
  units of $R_{2500}$. In both panels, dotted lines are the profiles for
  each single simulated cluster, while the heavy solid line is for the
  average profile. For reference, the vertical dashed line in the left
panel indicates the average value of $R_{2500}$.}
\label{fi:tproj} 
\end{figure*}

Observational data from spatially resolved spectroscopy with the ASCA
(e.g., Markevitch et al. 1998), Beppo--SAX (De Grandi \& Molendi 2002)
and XMM--Newton (e.g., Pratt \& Arnaud 2002) satellites show that temperature
profiles decline at cluster--centric distances larger than about one
quarter of the virial radius (see the data points in the left panel of
Figure \ref{fi:tproj}; cf. also Irwin \& Bregman 2000). Furthermore,
Beppo--SAX (De Grandi \& Molendi 2002; Ettori et al. 2002a), Chandra
(Ettori et al. 2002b; Allen, Schmidt \& Fabian 2001; Johnstone et
al. 2002) and XMM--Newton (Tamura et al. 2001) data show that temperature
profiles smoothly decline towards the cluster center. In particular,
Allen et al. (2001, A01 hereafter) analysed Chandra data for 6 fairly
relaxed hot clusters, with $T\magcir 5.5$ keV. They found profiles
which are quite similar once they are rescaled to $R_{2500}$ (the
radius encompassing an average density $\bar\rho/\rho_{\rm
crit}=2500$): an isothermal profile in the range $0.3\mincir
/R_{2500}\mincir 1$, with a smooth decline at smaller radii (dashed
curve in the right panel of Fig.~\ref{fi:tproj}).

While a declining profile in the outer cluster regions is generally
expected from hydrodynamical simulations (e.g., Evrard et al. 1996; Eke
et al. 1998; Borgani et al. 2002; Loken et al. 2002; Rasia et
al. 2003; Ascasibar et al. 2003), the
presence of a nearly isothermal core with a central smooth temperature
drop represents a significant challenge. In fact, including cooling
and star formation in simulations has the effect of steepening the
central temperature profile (e.g., Lewis et al. 2000; Muanwong et
al. 2002; Valdarnini 2003), opposite to the
observed decline of temperature. This is because cooling generates a
lack of pressure support in central regions, which causes gas infall
from outer regions, and this gas is heated by adiabatic compression as
it streams towards the centre (e.g., Tornatore et al. 2003).  A
possible remedy of this disagreement may lie in extra heating of the
ICM: placing the gas on a higher adiabat should reduce both the amount
of cooling and the compressional heating. However, Tornatore et
al. (2003) demonstrated that it is not easy to obtain this effect,
even when a variety of heating prescriptions as in their study are
explored.

\begin{figure*}
\centerline{
\hbox{
\psfig{file=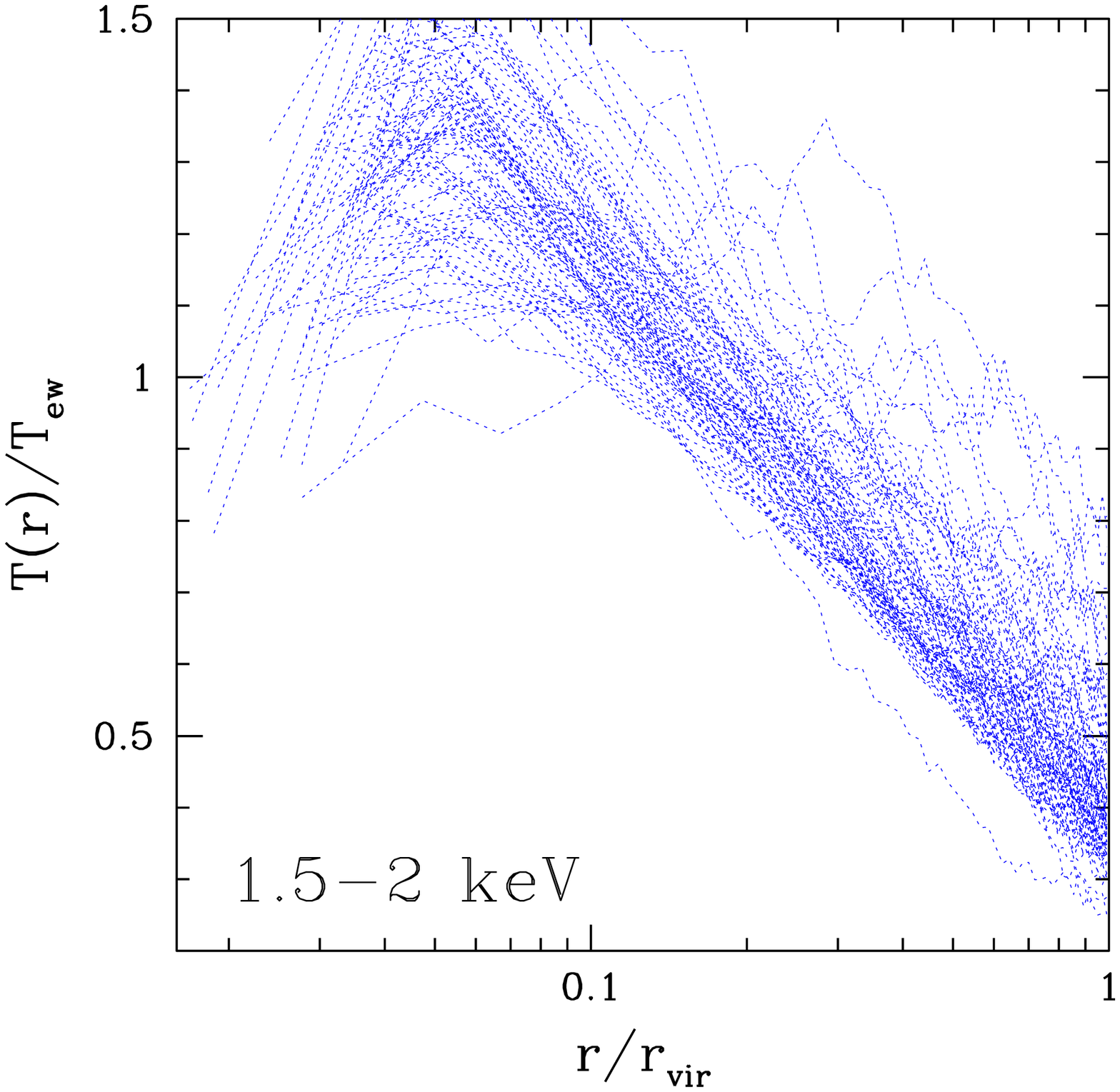,width=5.5cm} 
\psfig{file=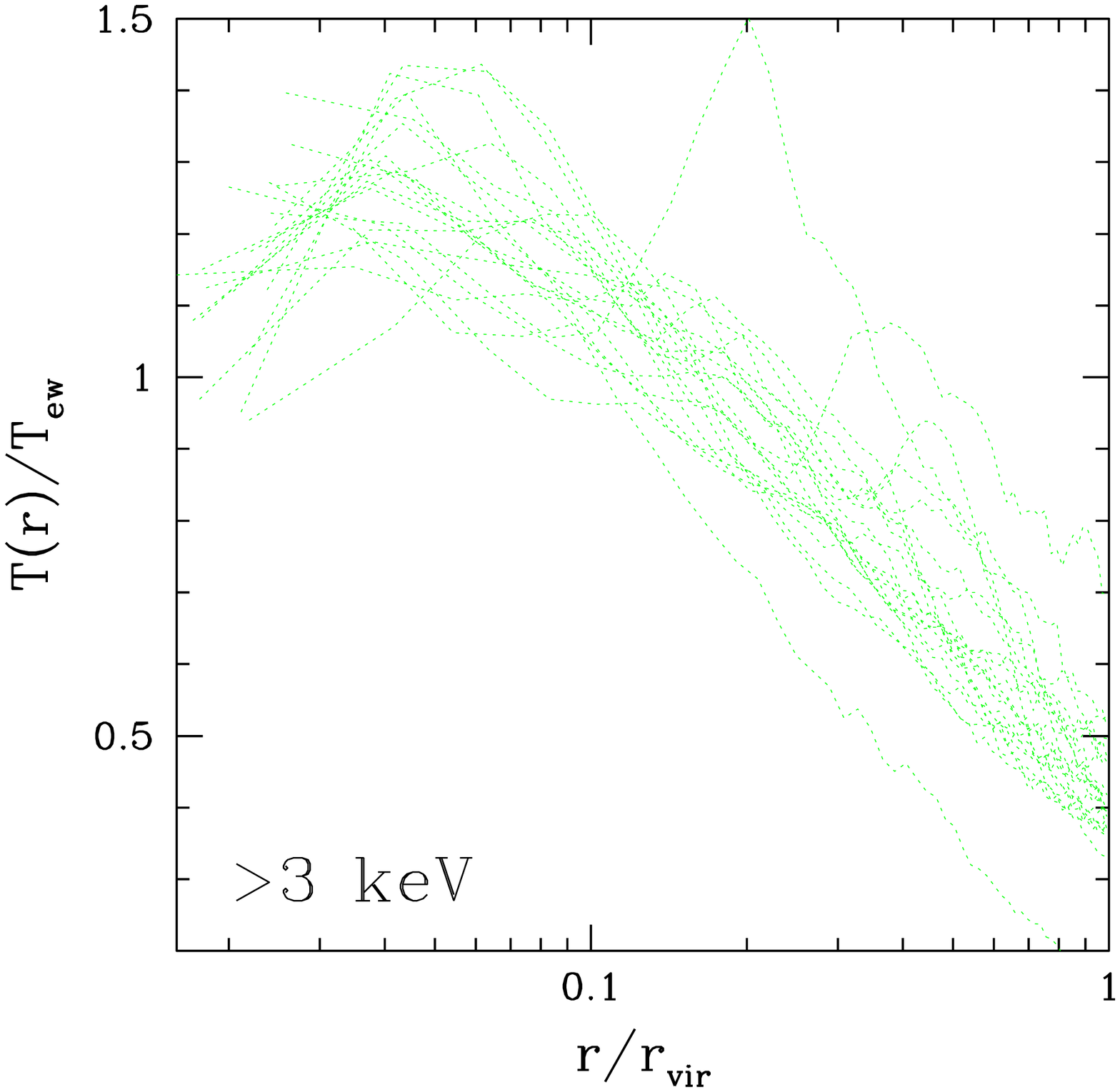,width=5.5cm} 
\psfig{file=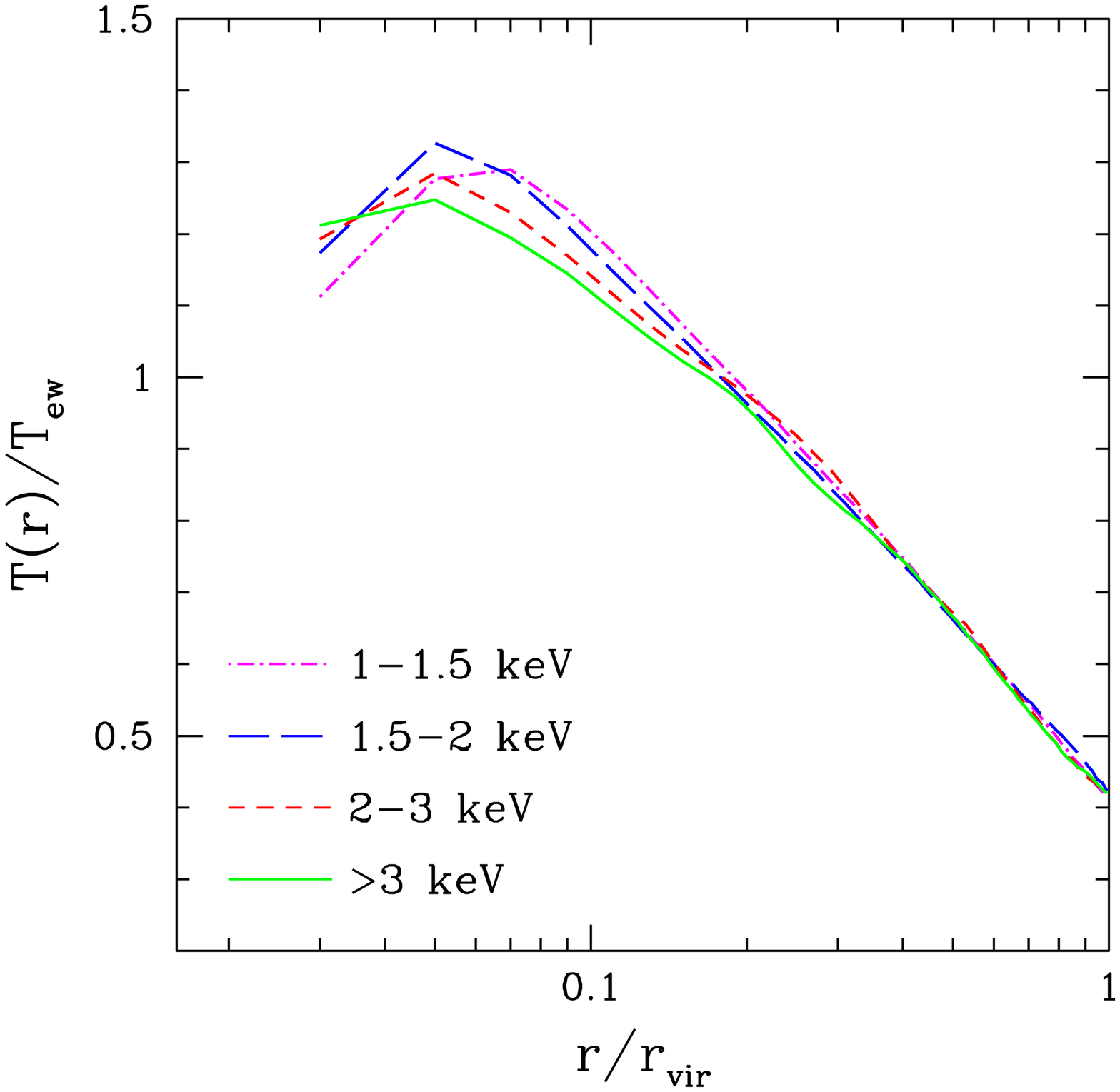,width=5.5cm} 
}}
\caption{The temperature profiles for objects of different
  temperatures. The left and the central panels show individual
  profiles, while the right panel shows the average profiles for
  groups and clusters within different interval of $T_{\rm ew}$.}
\label{fi:tprof} 
\end{figure*}

In Figure \ref{fi:tprof}, we show the three--dimensional temperature
profiles for our clusters and groups within different intervals of
$T_{\rm ew}$. Although there is a considerable scatter among these
profiles, they are all increasing with similar slopes down to $\simeq
0.05\,r_{\rm vir}$, with a temperature drop taking place only in the
innermost regions. Individual profiles, rather than being smooth, are
characterized by wiggles which are associated either with merging
sub--groups that contain relatively cold gas, or with supersonic
accretion which heats the gas across the shock fronts. These effects
cause the complex temperature structures detected in real clusters
(e.g., Mazzotta et al. 2002; Markevitch et al. 2003), and they are now
routinely reproduced in high-resolution simulations (e.g., Bialek,
Evrard \& Mohr 2002; Motl et al. 2003; Tormen, Moscardini \& Yoshida
2003).

In the right panel of Figure \ref{fi:tprof}, we plot the average
profiles for systems of different temperatures. Clusters and groups
follow a nearly universal temperature--profile, with only a marginal
tendency for hot systems to have shallower profiles at $r\mincir
0.2\,r_{\rm vir}$. In general, these profiles show neither evidence
for an isothermal core, nor a central smooth decline down to about 1/2
of the virial temperature, as it is observed in many real clusters.

A more direct comparison with observations is given in
Figure~\ref{fi:tproj}, where we compute projected profiles for the
simulated clusters. For a proper comparison with the results by De
Grandi \& Molendi (2002), we select only clusters with $T> 3$ keV,
which is the range covered by their 17 clusters observed with
Beppo--SAX. For each of the 23 simulated clusters with $T>3$ keV thus
selected, we plot the profiles projected along three orthogonal
directions. After projection, the average profile of the simulated
clusters does still not show evidence for an isothermal core.  They
steadely increase toward the cluster center down to $R\magcir
0.04R_{180}$ (we define $R_{180}$ as the radius at which
$\bar\rho/\rho_{\rm cr}=180$), while a temperature decrease is
observed only in the innermost regions. It is worth noting that the
slope of the simulated profile in the outer regions is similar to,
although slightly shallower than, the observed one, thus in agreement
with the ``universal'' temperature profile that Loken et al. (2002)
obtained for simulations without cooling.

To further demonstrate the failure of the simulation to account for
the observed central temperature profiles, we compare our results in
the right panel of Fig.~\ref{fi:tproj} to those obtained by A01 for a
set of six fairly relaxed clusters with temperatures $T_{2500}$ in the
range 5.5 to about 15 keV. Since in this range we have only one
cluster, with $T\simeq 7$ keV, we also use our five clusters with $T>
4$ keV for the comparison. This should not introduce any systematics,
at least as long as hot clusters are self--similar, which is actually
one of the claims made by A01. X--ray spectroscopy at high spatial
resolution with Chandra opened the possibility to trace the
temperature structure of such clusters down to unprecedented small
scales, which are, however, easily accessible by our simulation. Again,
the observed universal profile by A01 largely deviates from the
simulated one. We point out that the profiles from the simulation show
considerably more scatter than those of real clusters, but this is
just due to the fact that we did not attempt to select relaxed
clusters only.  Nevertheless, we emphasize that in no case we find a
cluster having an isothermal region followed by a smooth decline at
$R\mincir 0.3\,R_{2500}$.

We argue that this discrepancy between simulated and observed
temperature profiles is strong evidence for the current lack of
self--consistent simulation models capable of explaining the thermal
structure of the ICM in the regime where radiative cooling and
feedback heating are highly important. We shall further discuss this
point in Section~\ref{s:disc}.

\subsection{The luminosity--temperature relation}

Earlier on, it has been recognized that the $L_X$--$T$ relation of
clusters provides evidence for a lack of self--similarity in the ICM
properties (e.g., Evrard \& Henry 1991; Kaiser 1991). Its shape at the
cluster scale, $T\magcir 2$ keV, is well described by $L_X\propto
T^\alpha$ with $\alpha \simeq 2.5$--3 (e.g., White et al. 1997; Xue \&
Wu 2000), with a possibly shallower slope that approaches the
self--similar expectation of $\alpha=2$ for the very hot systems
(Allen \& Fabian 1998), and a considerably reduced scatter once
cooling flow clusters are removed (Arnaud \& Evrard 1999; Ettori et
al. 2002a), or when the contribution from cooling regions is excised
(e.g., Markevitch 1998).  Furthermore, evidence has been found that
groups with $T_X< 2\,{\rm keV}$ have a significantly steeper slope of
$\alpha \sim 5$ (e.g., Sanderson et al. 2003, and references therein),
although this result is not confirmed by the analyses by Mulchaey
\& Zabludoff (1998) and Osmond \& Ponman (2003).

\begin{figure}
\centerline{
\psfig{file=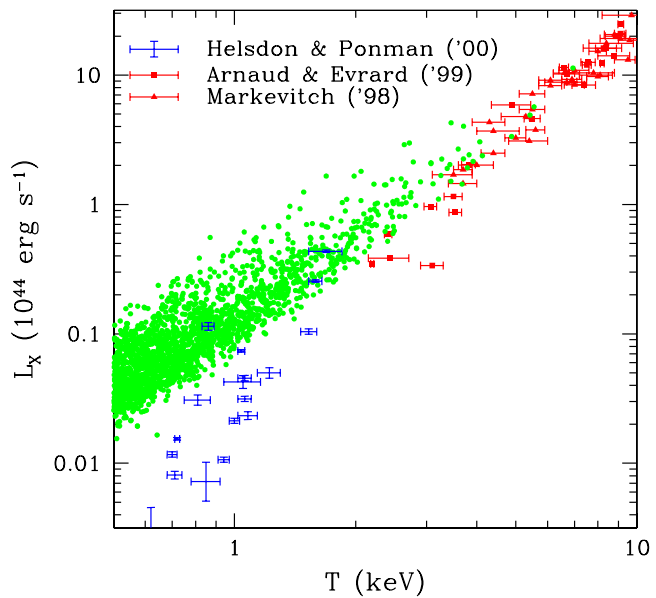,width=8.5cm} 
}
\vspace{-0.truecm}
\caption{Comparison between the observed and the simulated
  relation between bolometric luminosity, $L_X$, and emission--weighted
  temperature, $T_{\rm ew}$. The latter has been computed weighting the
  contribution from each gas particle according to its emissivity in
  the [0.5-10] keV energy band.}
\label{fi:lt} 
\end{figure}

In Fig.~\ref{fi:lt}, we show a comparison between the observed and
simulated $L_X$--$T$ relations for clusters and groups. Quite
apparently, the simulation results reproduce the observations
reasonably well on cluster scale. A log--log least--square fit to the
relation 
\be \log\left({L_X\over
L_{X,0}}\right)\,=\,\alpha\,\log\left({T_{500}\over {\rm keV}}\right)
\label{eq:ltfit}
\ee 
for clusters with $T>2$ keV gives $\alpha=2.5\pm 0.1$ and
$L_{X,0}=(1.0\pm 0.3)\,10^{43}\lum$, with intrinsic scatter ${\Delta
T\over T}=0.33$. Including colder systems in the fit, down to $T_{\rm
ew}=0.7$ keV, confirms the visual impression that no significant
change of slope takes place in the simulation at the scale of groups,
thus consistent with the result found by Muanwong et al. (2002).
While this is in contradiction with observational claims for a
steepening on group scales (e.g., Ponman et al. 1996; Helsdon \&
Ponman 2000; Sanderson et al. 2003), it agrees with other analyses
which indicate a unique slope from the cluster to the group scales
(Mulchaey \& Zabludoff 1998). Osmond \& Ponman (2003) have
recently analysed ROSAT--PSPC data for an extended set of galaxy
groups and, although within a large scatter, found no evidence for a
steepening of the $L_X$--$T$ relation.

From an observational point of view, determining the X--ray luminosity
contributed by the diffuse medium in galaxy groups is not as
straightforward as for richer clusters, mainly due to the
uncertainties in removing the contribution from member galaxies,
especially from the dominant ellipticals. For instance, if genuine
emission from the diffuse intra--group medium is removed when excising
galaxies, then the X--ray luminosity may be underestimated, thus
leading to a steepening of the $L_X$--$T$ relation. This point was
quite critical when considering pre--Chandra X--ray imaging, due to
the limited spatial resolution. However, there is no doubt that, as
Chandra data for a critical number of groups accumulates, it will
become possible to settle the question of how much of the X--ray
emission in central group regions has to be assigned to the diffuse
medium.

Quite interestingly, the intrinsic scatter in the simulated $L_X$--$T$
relation is rather small, comparable to the 30 per cent value reported
by Arnaud \& Evrard (1999), even though we did not attempt to correct
for the contribution of cooling regions. This is consistent with our
finding from Fig.~\ref{fi:lprof} where we do not detect significant
spikes of emissivity associated with central cooling regions. In order
to examine this further, we decided to apply to our clusters with
$T_X>2\,{\rm keV}$ the procedure adopted by Markevitch et al. (1998)
for excising the contribution from cooling regions. We first masked
regions smaller than 50 $h^{-1}$ kpc and then multiplied the resulting
$L_X$ by a factor 1.06 to account for the flux inside the masked
regions, thereby assuming a $\beta$--model with $\beta_{\rm fit}=0.6$
and core radius of 125 $h^{-1}$ kpc. As a result, we find that this
procedure affects the $L_X$ values only marginally, thus having only a
negligible effect on the scatter of the $L_X$--$T$ relation.

\subsection{The mass--temperature relation}

It is a well known problem of hydrodynamical simulations of cluster
formation with gravitational heating only that they predict a
normalization of the relation between total self--gravitating mass and
ICM temperature which is about 40 per cent higher than observed (e.g.,
Evrard et al. 1996).  This means that, for a fixed mass,
simulated clusters tend to be colder than observed (e.g., Horner,
Mushotzky \& Scharf 1999; Nevalainen, Markevitch \& Forman 2000; F01;
A01; Ettori et al. 2002a; Sanderson et al. 2003). On the scale of
groups, the slope of the mass--temperature relation is possibly also
steeper than the $M\propto T^{3/2}$ predicted
by hydrostatic equilibrium.

A possible interpretation for this discrepancy is that
non--gravitational heating increases the ICM temperature at a fixed
mass, thus lowering the $M$--$T$ amplitude. However, hydrodynamical
simulations that include non--gravitational heating have demonstrated
that this leaves only a negligible imprint in the $M$--$T$ relation:
after being preheated, the ICM settles back into hydrostatic
equilibrium, with its temperature being determined by the
DM--dominated gravitational potential well (Borgani et
al. 2002). Another possible explanation relies on the effect of
radiative cooling (e.g., Thomas et al. 2001; Voit et
al. 2002). Although this may appear counterintuitive, radiative
cooling may actually increase the ICM temperature, because
it eliminates central pressure support, thus
causing gas from outer regions to flow in and be heated by adiabatic
compression. Although simulations do show such an effect to some extent,
it is not clear whether cooling alone is able to reconcile the simulated
and observed mass--temperature relations (e.g., Muanwong et al. 2002; Tornatore
et al. 2003). Furthermore, for $T_{\rm ew}$ to increase by the required
amount, one is forced to increase the ICM temperature in central
regions, thus steepening the temperature profiles. As already
discussed this is not a welcome feature.

\begin{figure*}
\centerline{
\hbox{
\psfig{file=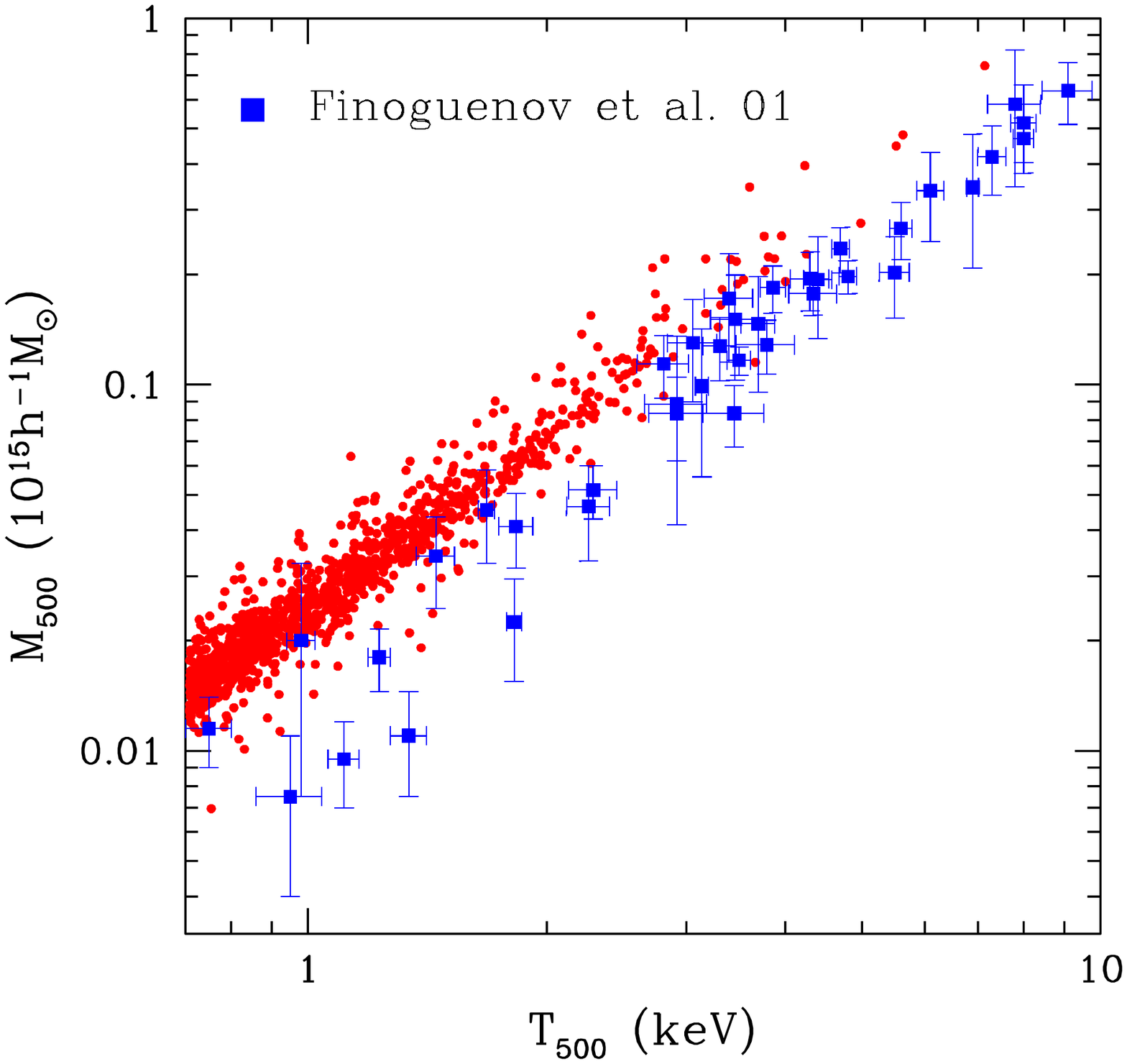,width=5.5cm} 
\psfig{file=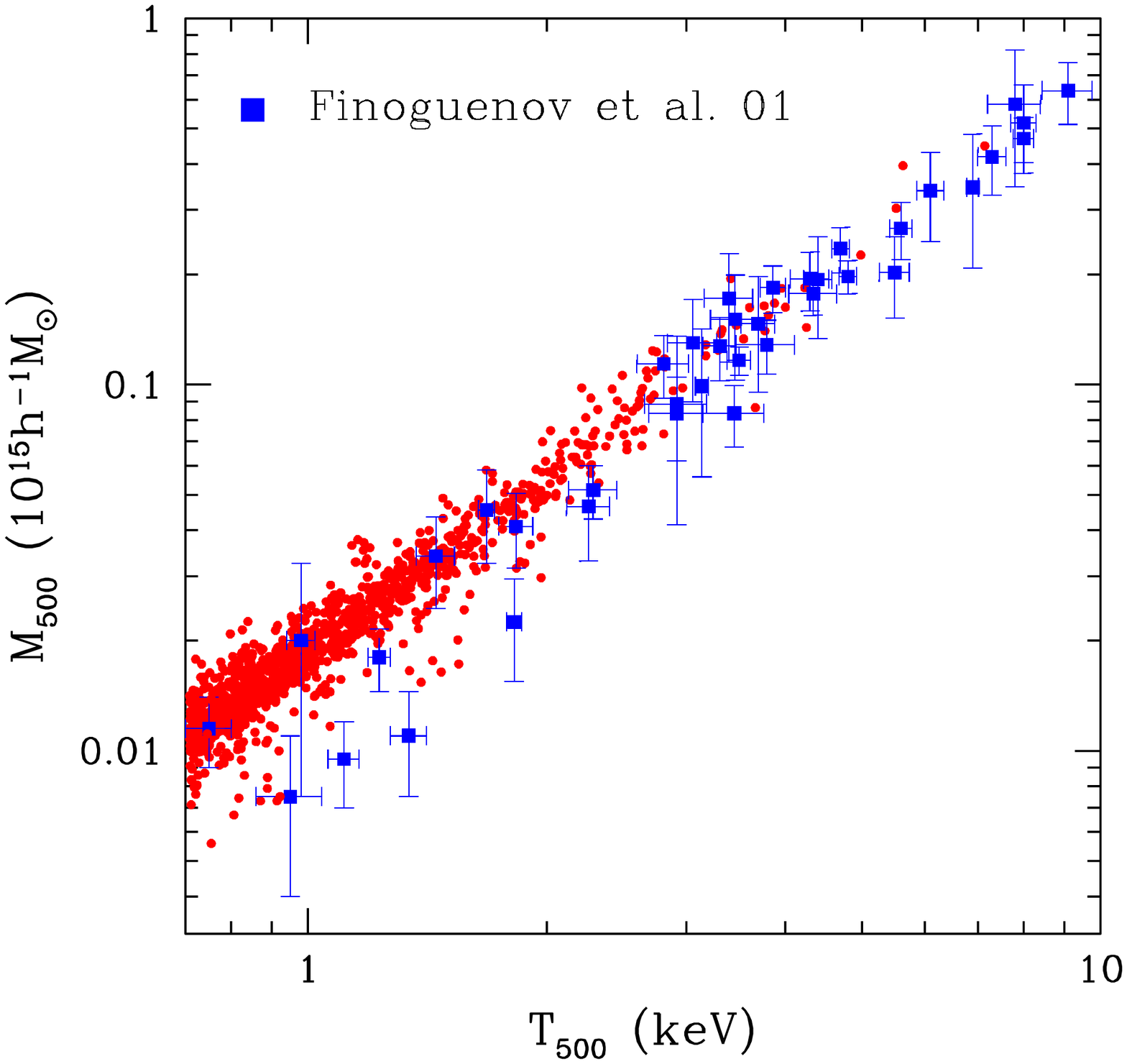,width=5.5cm} 
\psfig{file=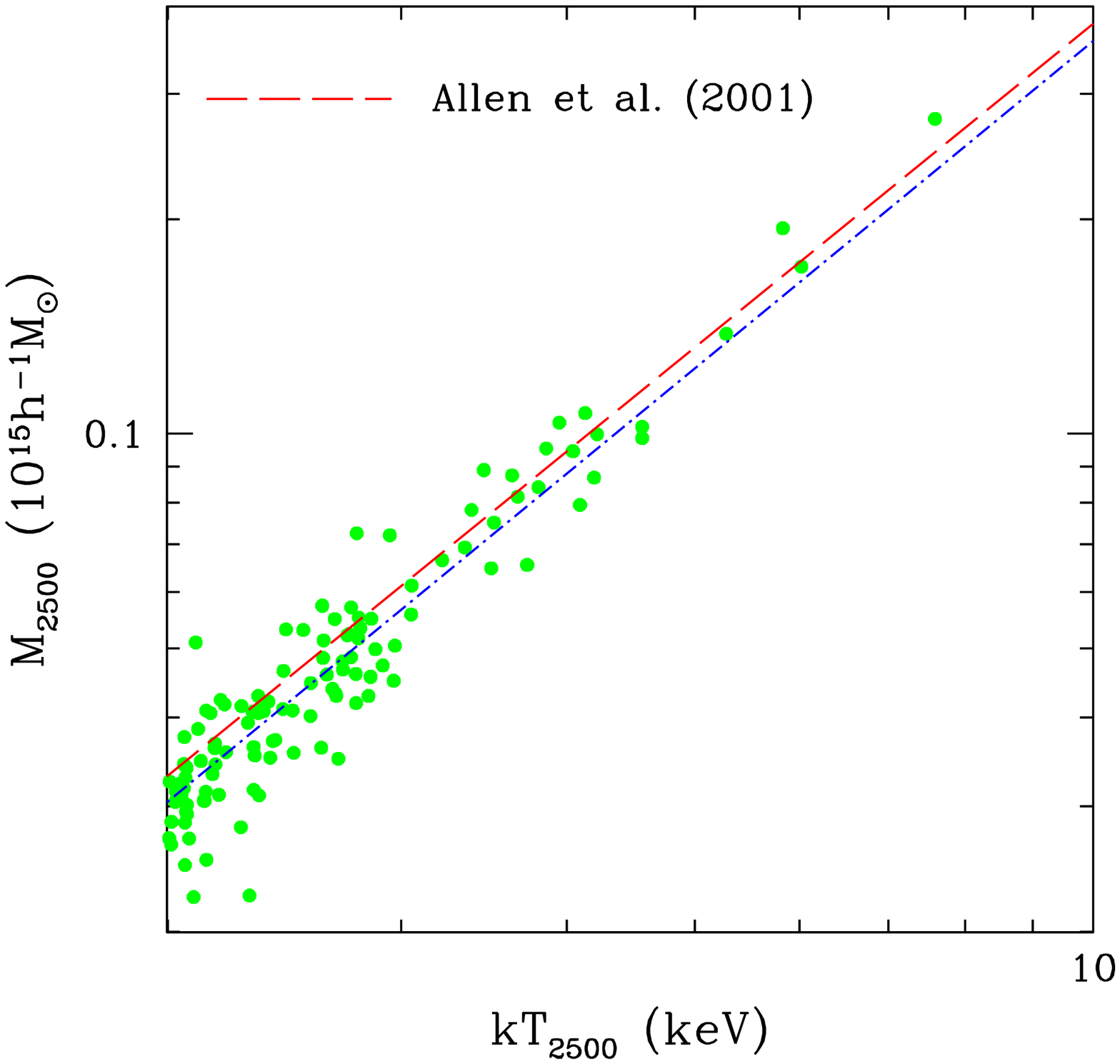,width=5.4cm} 
}}
\caption{Comparison between the observed and the simulated $M$--$T$
relation. Left and central panels refer to the relation at
$\bar\rho/\rho_c=500$. In the left panel we compare the results from
Finoguenov et al. (2001, F01) to the true total masses of simulated
clusters. In the central panel, cluster masses are estimated by
reproducing the procedure followed by F01 (see text): the equation of
hydrostatic equilibrium is applied after fitting the gas density
profile to a $\beta$--model and assuming a polytropic equation of
state. In the right panel the simulation results at
$\bar\rho/\rho_c=2500$ (points) are compared to the relation found by
Allen et al. (2001, A01) from Chandra for hot relaxed clusters (dashed
line); the dot--dashed line is our best--fit to the clusters with
$T_{2500}>2$ keV; we note that in this case $T_{2500}$ should be
interpreted as a mass--weighted temperature (see text).}
\label{fi:mt} 
\end{figure*}

The left panel of Figure~\ref{fi:mt} actually demonstrates that
cooling in our simulation is not effective in reducing the $M$--$T$
normalization to the observed level. In this panel, our results are
compared to those by F01. Making a log--log least
square fitting to the relation
\be
\log\left({M_{500}\over M_0}\right)\,=\,\alpha\,\log\left({T_{500}\over
    {\rm keV}}\right),
\label{eq:mtfit}
\ee 
we obtain $\alpha=1.59\pm 0.05$ and $M_0=(2.5\pm
0.1)\,10^{13}h^{-1}M_\odot$. Therefore, the normalization of our
relation at 1 keV turns out to be higher by about 20 per cent than
that found by F01, while it is lower by about 20 per cent with respect
to that found by Evrard et al. (1996) from simulations not including
radiative cooling.  The intrinsic scatter around the best--fitting
relation is $\Delta M/M=0.16$. Since simulation data have no
measurement error, this has to be interpreted as the intrinsic scatter
which originates from cluster dynamics.

A critical issue in this comparison concerns the different procedures
used for estimating masses in simulations and in observations. For
instance, Markevitch et al. (1998), Nevalainen et al. (2000) and F01
estimate masses by applying the equation of hydrostatic equilibrium,
assuming a $\beta$--model for the gas density profile and a polytropic
equation of state of the form $T\propto \rho_{\rm gas}^{\gamma-1}$,
where $\gamma$ is an effective polytropic index. Under these
assumptions, hydrostatic equilibrium provides the total
self--gravitating mass within the radius $r$ as
\be M(<r)\,=\,1.11\times 10^{14} \beta_{\rm fit}\gamma {T(r)\over {\rm
keV}} {r\over \hm} {x^2\over 1+x^2}\,,
\label{eq:hyeq}
\ee 
where $T(r)$ is the temperature at the radius $r$, $x=r/r_c$ is a
scaled radial coordinate in units of the 
core radius $r_c$ of the gas density profile, and we assumed
$\mu=0.6$ for the mean molecular weight. Based on hydrodynamical
cluster simulations, Bartelmann \& Steinmetz (1996) questioned this
procedure and suggested that the limited range of scales where the
$\beta$--model provides a good fit may bias cluster mass estimates
low by as much as 40 per cent (see also Muanwong et al. 2002).

\begin{figure}
\centerline{
\psfig{file=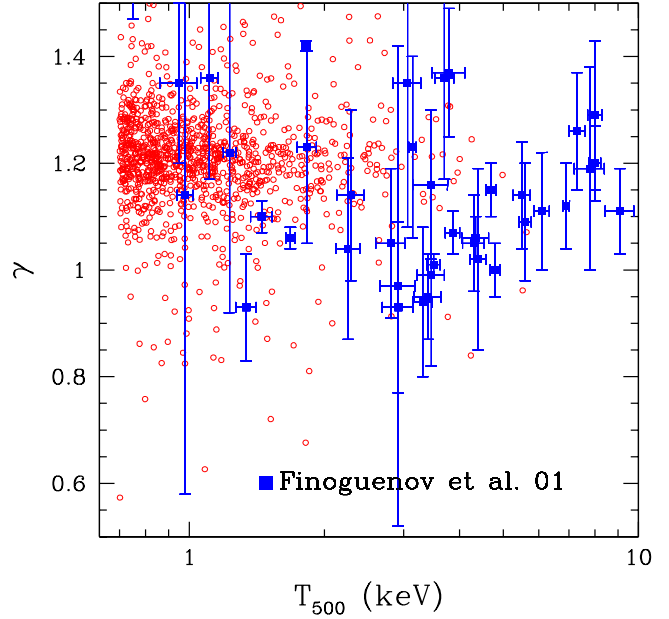,width=8.5cm} 
}
\caption{Comparison between the simulated (open circles) and the
  observed (filled squares, from Finoguenov et al. 2001) values
  of the effective polytropic index $\gamma$.}
\label{fi:gamma} 
\end{figure}

In order to verify this, we estimate cluster masses by applying
Eq.~(\ref{eq:hyeq}). For each cluster, we compute $\beta_{\rm fit}$ by
fitting the gas density profile to the $\beta$--model (see
Fig.~\ref{fi:bfit}), and $\gamma$ by fitting temperature and
gas--density profiles to the polytropic equation of state over the
same range of radii. The resulting values of $\gamma$ are compared in
Figure \ref{fi:gamma} to those reported by F01. Despite the fairly
large observational uncertainties, simulation results generally agree
with observations, with $\gamma\simeq 1.2$ and no dependence on
cluster temperature.  As for the core radius, both theoretical
arguments (e.g., Komatsu \& Seljak 2001) and observational data (e.g.,
F01) indicate that it is generally of the order of one--tenth of
$r_{500}$ so that the correction term in Eq.~(\ref{eq:hyeq}) for the
presence of a finite core is negligible.

As shown in the central panel of Fig.~\ref{fi:mt}, the effect of using
Eq.~(\ref{eq:hyeq}) as a mass estimator is that of lowering the
normalization of the $M$--$T$ relation and to bring it into better
agreement with observations.  While the slope and the intrinsic
scatter are left essentially unchanged, with $\alpha=1.57\pm 0.02$ and
$\Delta M/M\simeq 0.17$, the normalization is decreased to
$M_0=(1.9\pm 0.1)\,10^{13}h^{-1}M_\odot$.  The main reason for the
biasing towards low mass estimates lies in the limited range of scales
used to fit the $\beta_{\rm fit}$ parameter. In fact, we have verified
that, if we extend the fit out to the virial radius, the average value
of $\beta_{\rm fit}$ increases by about 15 per cent and, therefore, so
does the mass estimate. This result indicates that possible
biases in the observational mass estimates, related to the assumptions
of $\beta$--model profile and hydrostatic equilibrium (see also the
discussion by Rasia et al. 2003, and Ascasibar et al. 2003), may be at
the origin of the difference between our simulated $M$--$T$ relation
and the observed one.

With the availability of data of better quality from Chandra and
XMM--Newton, one can avoid some of the assumptions that enter the
derivation of Eq.~(\ref{eq:hyeq}).  Based on Chandra observations, A01
realized high--resolution imaging and spatially resolved spectroscopy
for six relaxed clusters with $T\magcir 5.5$ keV. Instead of assuming
a $\beta$--model and a polytropic equation of state, they applied a
deprojection technique to temperature and surface brightness profiles
(e.g., White et al. 1997) in order to reconstruct gas mass and total
mass profiles, assuming that the latter can be parametrized with a NFW
model (Navarro, Frenk \& White 1997). As such, the $M$--$T$ relation
obtained from their analysis can be directly compared to the
simulation result based on the ``true'' cluster masses.  Given the
relatively small field-of-view of the ACIS-S Chandra detector, they
however had to restrict their analysis to $R_{2500}$.  The resulting
best--fitting $M_{2500}$--$T_{2500}$ relation is plotted as a dashed
line in the right panel of Fig.~\ref{fi:mt} and compared to the
results of our simulation. Thanks to the possibility of resolving
temperature profiles, the values of $T_{2500}$ provided by A01 are
computed by mass--weighting the temperature determinations in
different radial bins. In order to reproduce the procedure followed by
A01, we compute $T_{2500}$ in simulated clusters by mass--weighting
the corresponding temperature profiles.  After fitting our results to
the power--law scaling of the $M_{2500}$--$T_{2500}$ relation for the
clusters with $T_{2500}> 2$ keV, we obtain $\alpha=1.55\pm 0.05$ and
$M_0=(1.0\pm 0.2)\,10^{13}h^{-1}M_\odot$.

It is quite remarkable that the simulation results now agree with
observations on the $M$--$T$ relation. Admittedly, given the small
number of hot clusters in our simulation, this comparison is based on
assuming that the best--fit relation by A01 can be extrapolated to
colder systems. Still, this result suggests that observed and
simulated $M$--$T$ relations agree with each other when high--quality
observational data are used which accurately resolves the
surface--brightness profile and the temperature structure of
clusters.

According to the above discussion, comparing the observed and the
simulated $M$--$T$ relation requires understanding how mass is
estimated from data. At the same time, it is also important to
understand how measured temperatures compare to the temperature
inferred from simulations. As we discussed in Section 3.2, including
the effect of metal lines could bias low the observed temperature,
thus affecting the $M$--$T$ relation (e.g., Mathiesen \& Evrard
2001). Accounting for these effect requires to self--consistently
trace the pattern of metal enrichment and to reproduce in detail the
observational setup. This will become mandatory as the level of precision
at which the X--ray properties of real and virtual clusters increases.

One of the most important applications of the calibration of the
$M$--$T$ relation and of its intrinsic scatter is to obtain
cosmological constraints from the cluster X--ray luminosity function
(XLF) and temperature function (XTF). Of particular interest is the
normalization of the power spectrum in terms of $\sigma_8$.  The
lower the normalization of the $M$--$T$ relation, the smaller the
mass of collapsed halos to be identified with clusters of a given
temperature, and consequently, the lower the normalization of the
power spectrum required for a given cosmological model to fit the
observed XTF (e.g., Ikebe et al. 2002; Seljak 2002; Pierpaoli et
al. 2003, and references therein). Huterer \& White (2002) have shown
that, to a good approximation, the values of $\Omega_m$ and $\sigma_8$
scale with the normalization $M_0$ of the mass--temperature relation
as 
$\Omega_m^{0.6}\sigma_8\propto M_0^{0.53}$. If we take the
result of our analysis that the observed $M_0$ may be biased low
by about 25 per cent at face value, we would then expect that the
infered $\sigma_8$ should be biased low by almost 15 per
cent. Furthermore, the value of the intrinsic scatter in the $M$--$T$
relation also affects the determination of $\sigma_8$ from the cluster
XLF and XTF (e.g., Borgani et al. 2001b; Pierpaoli et al. 2003). In
the presence of a significant scatter, the model--predicted
distribution functions are given by the convolution of the
cosmological mass function with a Gaussian function with width equal
to this scatter. Therefore, the larger the scatter, the lower the
power spectrum normalization required to fit the observed XLF or XTF.

Thanks to the good statistics offered by our simulation, a reliable
calibration of the scatter expected in the $M$--$T$ relation can be
obtained. We suggest that this calibration should be used for the
determination of cosmological parameters from the XTF and XLF. This
shows the importance of a precise calibration of the relations between
theory--predicted mass and X--ray observed quantities (e.g. Rosati,
Borgani \& Norman 2002, and references therein), and highlights the
important role that large cosmological simulations can play in
understanding the systematics involved (e.g., Viana et al. 2003).

\begin{figure}
\centerline{
\psfig{file=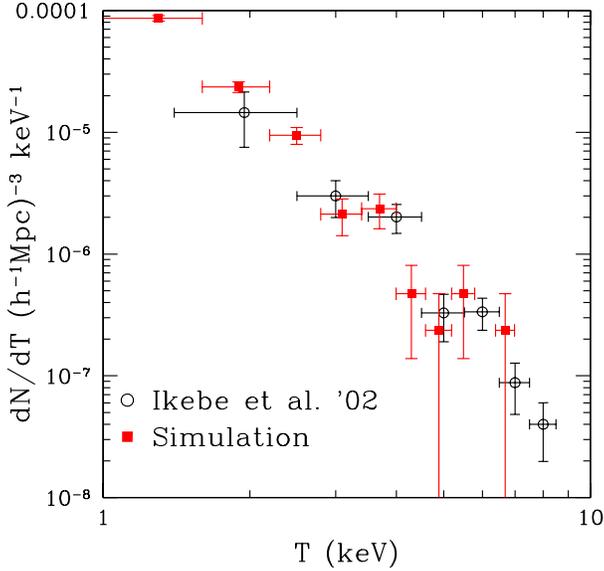,width=8.5cm} 
}
\vspace{-.5truecm}
\caption{Comparison between the simulated XTF at $z=0$ (filled
  squares) and the local XTF measured for the sample by Ikebe et al.
  (2002), and adapted for the simulated cosmological model (open
  circles; Ikebe, private communication). Error bars in the simulated
  XTF correspond to Poissonian uncertainties within each temperature
  bin.}
\label{fi:xtf} 
\end{figure}

In order to verify whether, with the chosen value of $\sigma_8=0.8$,
the simulation produces the correct cluster number density, we compare
in Figure \ref{fi:xtf} the XTF from our clusters to that reported by
Ikebe et al. (2002), adapted to our cosmological model (Ikebe, private
communication). Although our limited box size makes the comparison
difficult for $T> 6\,{\rm keV}$, we note that our simulated XTF agrees
with the observed one remarkably well over the common temperature
range. Since there are no adjustable parameters in the estimate of our
XTF, this indicates that $\sigma_8\simeq 0.8$ (for a flat model with
$\Omega_m=0.3$) is in fact required to match the observed number
density of nearby clusters.

\subsection{The $M_{\rm gas}$--$T$ relation}
The mass of diffuse gas is a quite useful diagnostic for the physical
status of the ICM, since it is free from several of the uncertainties
which affect the estimate of the total self--gravitating mass. Under
the assumption that gas follows dark matter, we expect that the gas
fraction in clusters, $f_{\rm gas}$ is independent of cluster mass
and, therefore, $M_{\rm gas}\propto T^{3/2}$.  Measurements of the gas
mass from X--ray observations have shown that the observed relation
between gas mass and temperature can indeed be represented by a power
law, $M_{\rm gas}\propto T^\alpha$, but with a slope which is
generally steeper than this self--similar expectation. For instance,
Vikhlinin et al. (1999) fitted gas density profiles out to $R_{200}$
and found $\alpha = 1.71\pm 0.13$. Other
estimates of the $M_{\rm gas}$--$T$ scaling at higher overdensity
find steeper slopes. Mohr et al. (1999) analysed a set of
clusters with $T\magcir 2$ keV at $\bar \rho /\rho_{cr}=500$ and found
$\alpha =1.98\pm 0.18$, while Ettori et al. (2002a) found
$\alpha=1.91\pm 0.29$ at $\bar \rho /\rho_{cr}=2500$ for a sample of
$T>3$ keV clusters. These results show that colder systems tend to be
less gas-rich, an effect which is more pronunced at higher
overdensity. These trends are naturally expected in scenarios where
self--similarity is broken by extra heating (e.g., Bialek et
al. 2001), which prevents gas from reaching high density in central
regions. Cooling can in principle also account for the observed
$M_{\rm gas}$--$T$ relation, provided its efficiency is significantly
higher in groups than in rich clusters.

\begin{figure}
\centerline{
\psfig{file=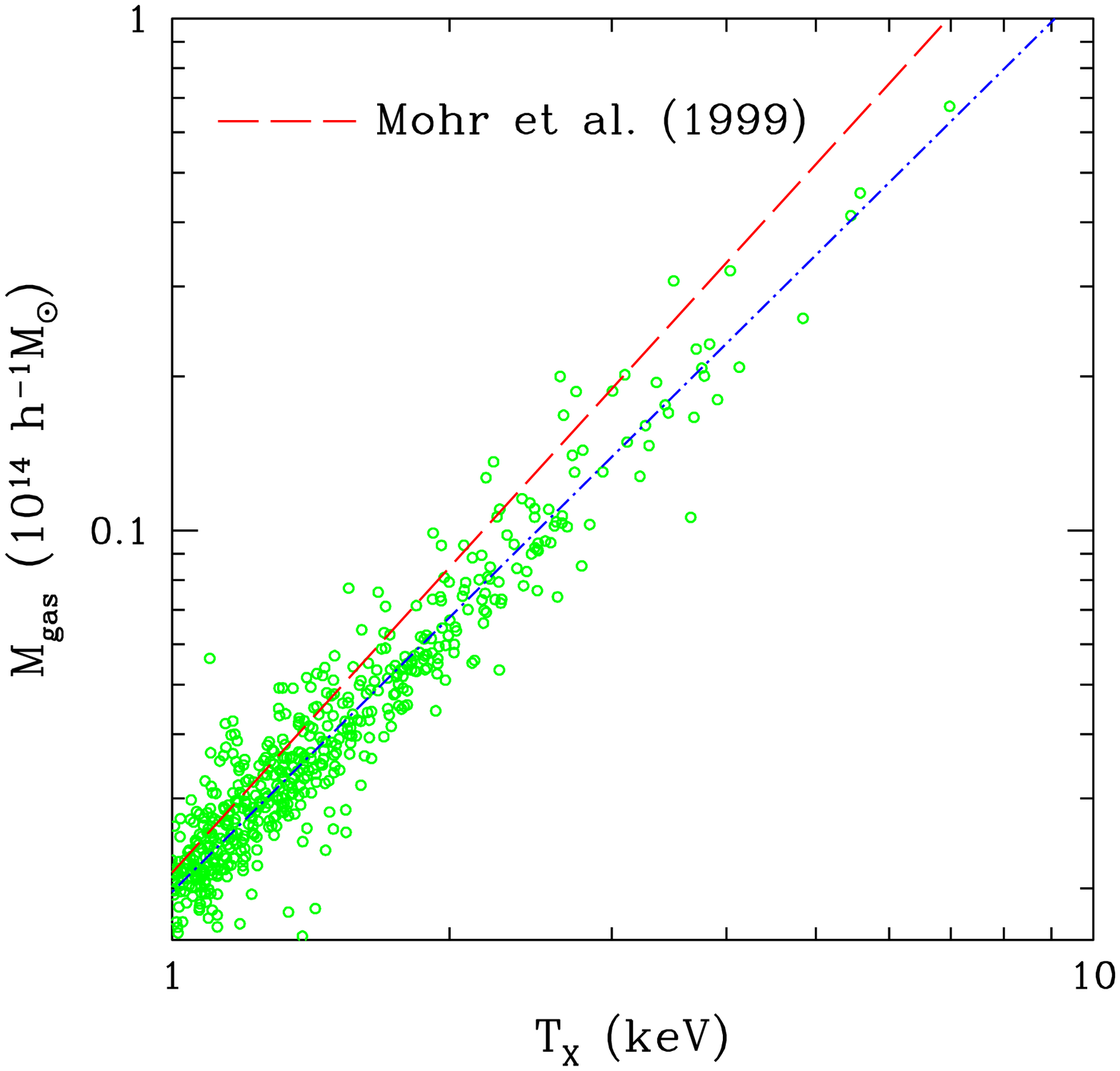,width=8.5cm} 
}
\caption{Comparison between the simulated (open circles) and observed
  $M_{\rm gas}$--$T$ relation at overdensity $\bar\rho
  /\rho_{cr}=500$.  The dashed line is the best--fitting model by Mohr
  et al. (1999) to their observational data points, $M_{\rm gas}=2.1
  (T_{\rm ew}/{\rm keV})^{1.98}10^{12}h^{-1}M_\odot$, while the
  dot-dashed line is the best--fitting relation to the results of the
  simulation $M_{\rm gas}=2.0 (T_{\rm ew}/{\rm
  keV})^{1.75}10^{12}h^{-1}M_\odot$.}
\label{fi:mgt}
\end{figure}

In order to compare results from our simulation with observations,
we make a log--log least--square fitting to the expression
\be
\log\left({M_{\rm gas}\over M_0}\right)\,=\,\alpha \,\log\left({T\over
    {\rm keV}}\right)\,.
\label{eq:mgfit}
\ee 
We find $\alpha=1.80\pm 0.08$ and $M_0=(1.8\pm
0.2)\,10^{12}h^{-1}M_\odot$ when fitted at $R_{500}$ for clusters with
$T_{500}> 2$ keV. As shown in Figure \ref{fi:mgt}, the simulated
relation tends to be shallower than the observed one, thus further
indicating that the feedback energy provided by our modelling of SN
explosions is not strong enough to break self--similarity as strongly
as observed. Note however that if a stronger heating was realized, the
effect would be that of suppressing $M_{\rm gas}$ more strongly
in low--mass systems, thereby reducing the overall normalization below
the observed level. The offset in the normalization is likely to be
due to two main reasons. First, since cooling in the simulation is too
efficient, too large a fraction of gas is removed from the X--ray
emitting phase. Therefore, reducing the cold phase to the observed
level would imply a compensating increase of $M_{\rm gas}$ by about 10
per cent. Second, our run is based on assuming $\Omega_{\rm bar}=0.04$
for the baryon density parameter. If the 20 per-cent larger value
indicated by WMAP data were used instead, this would have led to a
corresponding increase of $M_{\rm gas}$ of similar size.

\begin{figure*}
\centerline{\vbox{
\hbox{
\psfig{file=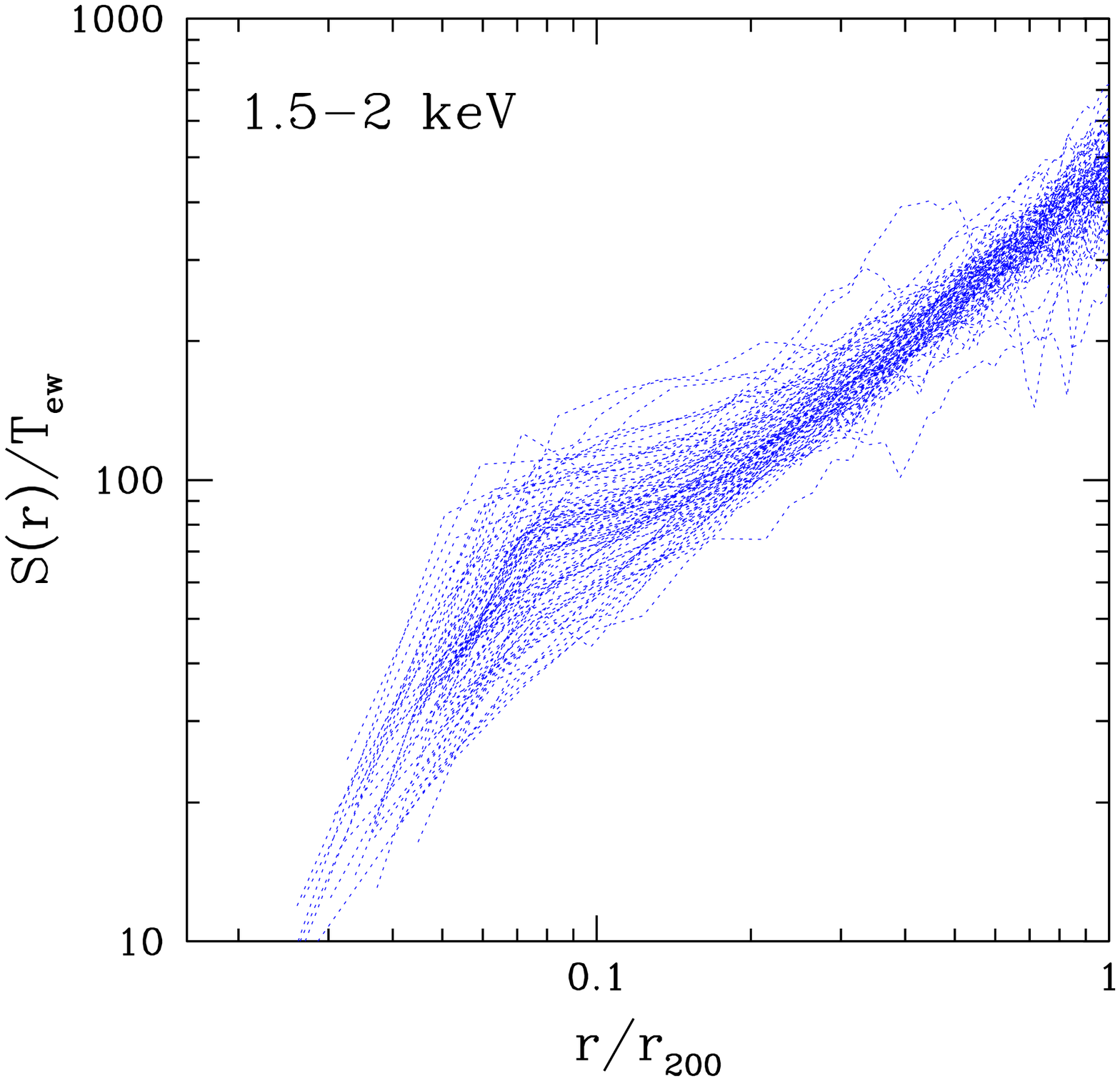,width=6.5cm} 
\psfig{file=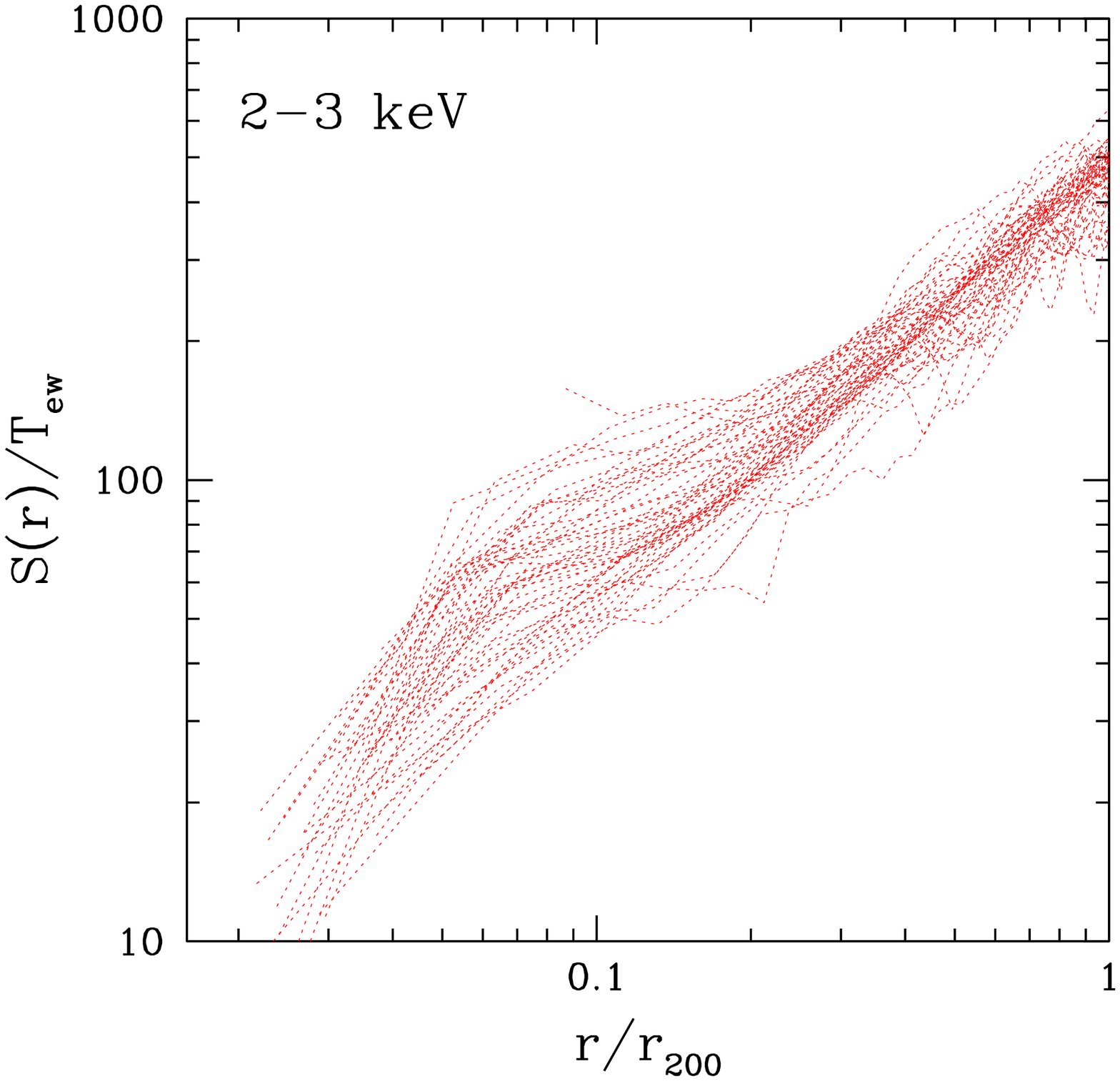,width=6.5cm} 
}
\hbox{
\psfig{file=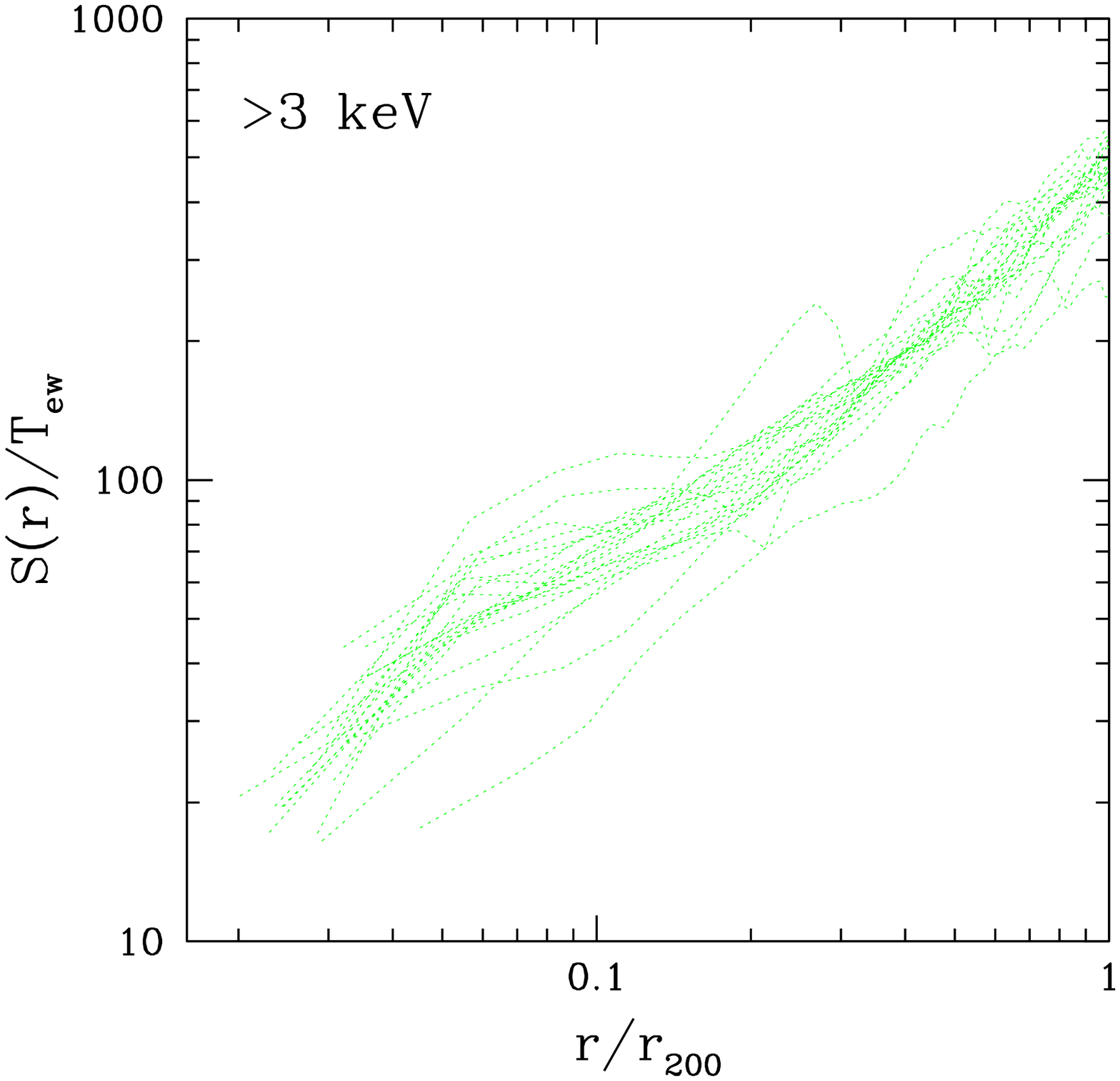,width=6.5cm} 
\psfig{file=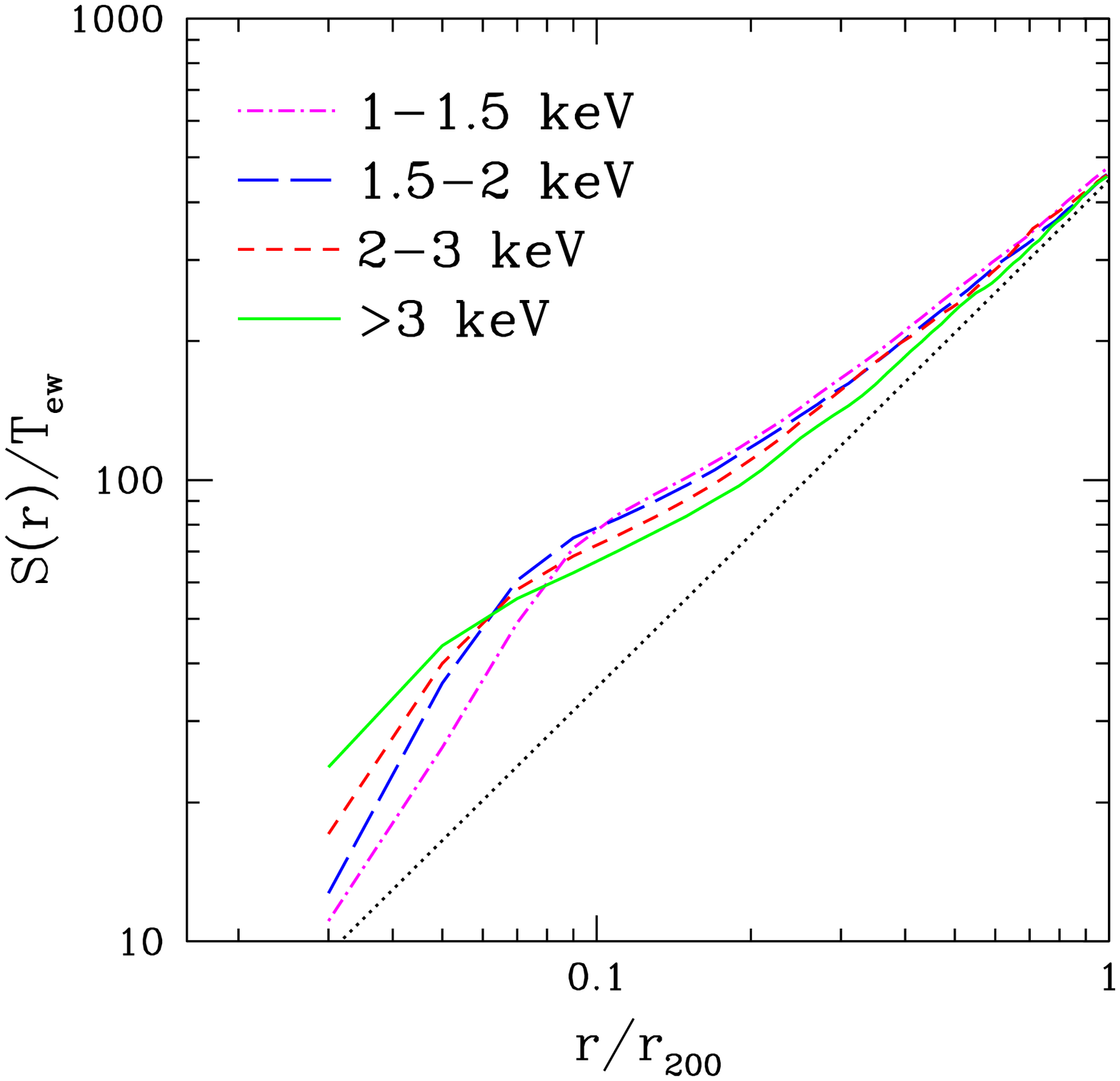,width=6.5cm} 
}
}}
\caption{Profiles of reduced entropy, $S/T$, at different intervals of
  $T_{\rm ew}$. The upper and the lower left panels show the profiles
  for all the structures within each temperature interval. The lower
  right panel shows the average profiles computed within each range of
  $T_{\rm ew}$.  In this panel the dotted line shows the scaling
  $S\propto r^{1.1}$, which is expected if entropy is generated by
  self--similar accretion shocks (Tozzi \& Norman 2001).}
\label{fi:sprof} 
\end{figure*}

\subsection{The entropy of the ICM}
Gas entropy is currently receiving considerable attention as a
diagnostic tool for tracing the past dynamical and thermal history of
the ICM (e.g., Bower 1997; Ponman et al. 1999; Tozzi \& Norman 2001;
Balogh et al. 2001; Borgani et al. 2001; Voit et al. 2002; Babul et
al. 2002).  In X--ray studies of galaxy clusters, the `entropy' is
commonly defined as
\be
S\,=\,{T\over n_e^{2/3}}\,.
\label{eq:ent} 
\ee
With this definition, $S$ is proportional to the adiabats of an ideal
monoatomic gas, and is expressed in units of ${\rm keV\, cm}^2$.  The
relation of $S$ to the thermodynamic definition of entropy, $s$, is
given by $s=c_V\log{S}+{\rm const}$, where $c_V$ is the specific heat
capacity at constant volume.

As long as the gas adiabat is not altered by some non--gravitational
heating source, one can derive the expected scaling between entropy
and temperature at redshift $z$ in the form $S\propto T^2/(E_z^2
\Delta_c)^{2/3}$, where $E_z=[(1+z)^3\Omega_m+1-\Omega_m]^{1/2}$
describes the evolution of the Hubble constant and $\Delta_c$ is the
overdensity, in units of the critical density, at which the entropy is
measured. This implies that, once $\Delta_c$ is fixed, entropy has a
linear dependence on gas temperature. In fact, when going from rich
clusters to poor groups, we probe gas that has been accreted at higher
redshift, when it was harder to generate accretion shocks and,
therefore, to increase the gas entropy.

\begin{figure}
\centerline{
\psfig{file=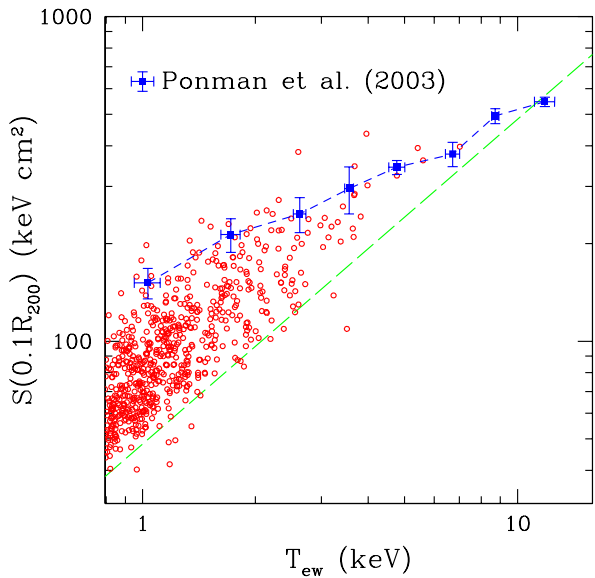,width=8.5cm} 
}
\vspace{-.5truecm}
\caption{The relation between gas entropy computed at $0.1\,r_{\rm
vir}$ and $T_{\rm ew}$. Data points are taken from Ponman et
al. (2003).  The dashed line shows for reference the self--similar
scaling $S\propto T$, normalized to the hottest cluster found in the
simulation.}
\label{fi:sthr} 
\end{figure}

For this reason, the entropy excess measured in the central regions of
poor clusters and groups (e.g., Ponman et al. 1999; Lloyd--Davies et
al. 2000; Finoguenov et al. 2002) is considered to provide direct
evidence that some non--gravitational process must have modified the
gas adiabat. For instance, if some process establishes a pre--collapse
entropy floor, we expect to measure the level of this floor at the
center of poor clusters and groups, while the scaling $S\propto T$
should be recovered for hot systems, whose entropy creation has been
dominated by gravitational shocks (e.g., Tozzi \& Norman
2001). However, this scenario is now questioned by the detection of
the so--called ``entropy ramp'' in central cluster regions (Ponman et
al. 2003; see the data points plotted in Figure \ref{fi:sthr}):
instead of following the $S\propto T$ scaling for hot systems with a
flattening to some floor for cold systems, entropy is shown to
gradually deviate from the self--simular scaling, with a temperature
dependence close to $S\propto T^{2/3}$.  Predictions of the standard
pre--heating scenario are also at variance with the results by
Mushotzky et al. (2003) from XMM--Newton observations of two groups,
which do no show any flattening of the entropy profiles in central
regions.

These results therefore call for alternative scenarios for gas
heating (e.g., Dos Santos \& Dor\'e 2002), or for a significant effect
of radiative cooling. Counterintuitively, cooling has the effect of
providing an increase of the entropy of the diffuse hot gas as a
result of selective removal of low--entropy gas from the hot phase
(e.g., Voit \& Bryan 2001; Voit et al. 2002; Wu \& Xue 2002). However,
while numerical simulations have shown that cooling and star formation
do indeed increase entropy in central cluster regions, they are not
able to account for the large excess observed on the scale of groups
(Finoguenov et al. 2003; Tornatore et al. 2003). Since these numerical
results were based on simulations of a small number of halos, they
were, however, suspicious of not being representative for the whole
population of galaxy systems. This limitation is overcome by the
fairly large set of clusters and groups available from our simulation.

In Figure \ref{fi:sprof}, we show the profiles of the ``reduced''
entropy, $S/T$. Note that these profiles should coincide if clusters
and groups are self--similar. Radii are given in units of $r_{200}$ so
as to make our profiles directly comparable to those recently
published by Ponman et al. (2003).  At large radii the profiles
approach the scaling $S(r)\propto r^{1.1}$, which is predicted in the
shock--dominated regime of entropy production by models based on
spherical gas accretion within a NFW dark matter halo (Tozzi \& Norman
2001; see the dotted curve in the lower right panel). This
demonstrates that gravity dominates the ICM thermodynamics in the
outer regions of clusters. Quite interestingly, this also indicates
that, although accretion is anisotropic in cosmological environments
where subgroups and gas primarily flow along filaments onto the
cluster (e.g. Tormen 1997), spherical accretion still provides an
adequate {\em average} description.

Looking at the average profiles (lower right panel) it is then
remarkable that they all fall on top of each other, almost independent
of the temperature of the system. Although this is quite expected, as
long as non--gravitational effects are negligible, it conflicts with
the recent results by Ponman et al. (2003). These authors computed
entropy profiles from ROSAT--PSPC and ASCA data for clusters and
groups in the temperature range 0.3--17 keV. Although the profiles
were found to be parallel to each other in the outer regions, and
close to the $r^{1.1}$ scaling, colder systems turned out to have a
relatively higher entropy. With an independent analysis, Pratt \&
Arnaud (2003) used XMM--Newton data for a cluster and a group, with
well resolved temperature profiles. They consistently found the
profiles of reduced entropy to be quite parallel to each other, with
the group having a relatively higher entropy, although intrinsic
scatter (see Fig. \ref{fi:sprof}) may limit the significance of a
result based only on two profiles.

Models based on an entropy floor or on cooling can account for the
lack of self--similarity in central cluster regions, while they
predict that all the systems follow self--similarity in regions close
to the last--shock radius.  Possible explanations for this global
violation of self--similarity have been proposed by Ponman et
al. (2003) and Voit et al. (2003), and are based on a differential
entropy amplification by shocks in the presence of
pre--heating. Smaller mass systems are expected to accrete from
lower--density filaments. Therefore, pre--heating should be more
efficient in evaporating condensations (Ponman et al.) and/or in
suppressing gas density (Voit et al.)  within the filaments accreting
into groups, compared with those accreting into clusters.  In turn,
this would make shocks stronger and, therefore, entropy generation
more efficient at the outskirts of colder systems. The fact that this
is not detected in our clusters may imply that any pre--heating
associated with SN energy feedback in our simulation is too week to
have a sizeable effect.

The entropy level at $r\simeq 0.1\,r_{200}$ is a factor $\sim 2$
larger than what is expected from the extrapolation of the $r^{1.1}$
scaling. At smaller radii, the entropy profiles tend to steepen, which
indicates the presence of a population of gas particles which start
feeling the effect of cooling, while still belonging to the hot
phase. This feature is more pronunced for smaller systems, whose
relatively higher central density makes cooling more efficient (as
also confirmed by their larger star fraction; see Fig.\ref{fi:fstar}).
Still, at $r\simeq 0.1r_{200}$ there are no significant differences in
the values of the reduced entropy. This is actually confirmed by
Figure \ref{fi:sthr}, where we compare the reduced entropy found for
simulated clusters at $0.1\,r_{200}$ to the ``entropy ramp'' measured
by Ponman et al. (2003). While our simulation confirms the existence
of an entropy excess, its magnitude does not depend sensitively on
temperature, in contrast to the observed gradual departure of entropy
from the expectation of self--similar scaling (dashed line) when
colder systems are considered. 

This result corroborates the finding that the adopted recipe for
supernova feedback from stellar populations, which includes
comparatively weak galactic winds, is not strong enough to increase
the gas adiabat in central cluster regions. At the same time,
radiative cooling is not able by itself to break the self--similarity
of the ICM entropy structure as strongly as observed.
Muanwong et al. (2002) found a better agreement of simulations with
the observational data by Ponman et al. (1999) with respect to the
entropy level in the central region of groups. Besides the difference
in the data set they are comparing to, the reason for this could also
be due to the fact that their simulation assumed a global gas
metallicity of $0.3\,Z_\odot$ at $z=0$ in the computation of the
cooling function (although metal production was not treated in their
code). A non--vanishing metallicity increases the cooling function
and, therefore, increases the entropy level for gas removal from the
hot phase (e.g., Voit \& Bryan 2001).

\section{Discussion and conclusions}
\label{s:disc}

We have presented results on the X--ray properties of groups and
clusters of galaxies, extracted from a large hydrodynamical simulation
within a cosmological box of $192\hm$ on a side. The simulation
includes radiative gas cooling, heating from a uniform UV background,
a description of star formation within a multi--phase interstellar
medium and a phenomenological treatment of galactic winds, powered by
supernova (SN) energy release. The quite high number of gas and DM
particles used, $2\times 480^3$, and the force resolution of 7.5
$h^{-1}$kpc provide high enough mass and force resolution for an
accurate description of the X--ray properties of galaxy systems down
to a virial temperature of about 0.5 keV. The results of this
simulation have been thoroughly compared to observational results on
the X--ray properties of clusters. Rather than focusing on quantities
that show good agreement between simulation and observations, this
comparison was made in the spirit of better understanding how our
current description of the physics of the ICM needs to be improved.

Our main results can be summarized as follows:
\begin{itemize}
\item[(a)] The scaling relations of X--ray luminosity and gas mass
with cluster temperature deviate from the predictions of self--similar
scaling. However, in some cases these deviations are not as large as
needed to agree with observations. For instance, the simulated
$L_X$--$T$ relation provides a good fit to data at $T\magcir 2$ keV,
while it does not produce any steepening at the scale of groups (e.g.,
Lloyd--Davies \& Ponman 2000; cf. also Mulchaey \& Zabludoff 1998). 
A recent analysis by Osmond \& Ponman (2003) actually demonstrates
  that the present quality of data is not good enough to allow a
  precise determination of the $L_X$--$T$ scaling for galaxy groups.
Furthermore, the $M_{\rm gas}$--$T$ relation is shallower than
observed (e.g., Mohr et al. 1999; Ettori et al. 2002a).
\item[(b)] Consistently, the gas density profiles are steeper than
observed, especially for groups. Fitting the profiles to a
$\beta$--model gives $\beta_{\rm fit}$ in the range 0.6--0.8, with no
appreciable dependence on temperature.
\item[(c)] The simulated $M$--$T$ relation is about 20 per cent higher
than that measured from observations under the assumptions of a
$\beta$--model in hydrostatic equilibrium with a polytropic equation
of state (e.g., Horner et al. 1999; Nevalainen et al. 2000; Finoguenov
et al. 2001). If masses of simulated clusters are estimated with this
same procedure, they are found to be biased low by just the amount
required to recover agreement with the observed $M$--$T$
relation. Quite interestingly, a good agreement is in fact found by
comparing simulation results to the $M_{2500}$--$T_{2500}$ relation
based on the Chandra data by Allen et al. (2001), which does not rely
on the assumption of a specific gas density profile. This suggests
that the problem of the $M$--$T$ normalization may be solved by a
better data treatment.
\item[(d)] The X--ray temperature function (XTF) from the simulation
agrees quite nicely with the most recent observational determinations
(e.g., Ikebe et al. 2002; Ikebe, private communication). This
indicates that the chosen power--spectrum normalization,
$\sigma_8=0.8$, for $\Omega_m=0.3$ is consistent with the measured
number density of galaxy clusters.
\item[(e)] Temperature profiles from the simulation are discrepant
with respect to observations. While their shape in the outer regions,
at $R\magcir 0.3 R_{180}$, is similar to the observed one, simulation
profiles are steadily increasing towards the centers, with no evidence
for an isothermal regime (e.g., De Grandi \& Molendi 2002) followed by
a smooth decline at $R\mincir 0.3\,R_{2500}$ (e.g., Allen et
al. 2001).
\item[(f)] The entropy properties of simulated clusters are also quite
different from observational results. In the outer regions, the
entropy profiles from the simulation are remarkably self--similar,
while observations show evidence for excess entropy at the scale of
groups (Ponman et al. 2003). In the inner regions, we detect a
significant excess entropy whose amount is almost independent of the
cluster temperature. Although the resulting $S$--$T$ relation
therefore deviates from the self--similar expectation, it is anyway
steeper than observed (Ponman et al. 2003).
\item[(g)] The fraction of baryons which cool and turn into stars
within the virial regions of clusters is $f_*\simeq 20$ per cent, with
a slight tendency to be higher for colder systems. This value is
substantially smaller than the one found in simulations that do not
include efficient feedback mechanisms, but it is still higher than
observed by about a factor of two (e.g., Balogh et al. 2001; Lin et
al. 2003). This demonstrates that the choice of SN feedback included
in the simulation is not efficient enough to prevent overcooling. We
note that in the simulations of Springel \& Hernquist (2003b) lower
stellar fractions were obtained, but these authors adopted galactic
winds that were twice as energetic as the ones included here.
\end{itemize}

In general, these results show that the physical processes included in
our simulation are able to account for the basic global properties of
clusters, such as the scaling relations between mass, temperature and
luminosity. At the same time, we find indications suggesting that a
more efficient way of providing non--gravitational heating from
feedback energy compared to what is implemented in the simulation is
required: this `extra heating' should not only reduce the amount of
gas that cools, but also needs to `soften' the gas density profiles of
poor clusters and groups by increasing the entropy of the relevant
gas. 
We also remind that the present simulation has been realized
using a zero--metallicity cooling function. Although the effect of
metal--line cooling is expected to marginally affect the overall
cosmic star formation (e.g., Hernquist \& Springel 2003), it is known
to increase quite significantly the X--ray emissivity of gas at
$T\mincir 2$ keV, as well as the efficiency in the removal of
low--entropy gas from the hot phase (e.g., Voit et al. 2002).

We think that it is unlikely that the problems of the present
simulation 
are caused by numerical limitations. The model for star formation and
galactic winds triggered by SN-II feedback, as implemented in our
simulation (Springel \& Hernquist 2003a), has been demonstrated to
have well-behaved numerical properties, and it produces a realistic
and numerically convergent cosmic star--formation history (Springel \&
Hernquist 2003b) that can be accurately understood by analytic
reasoning (Hernquist \& Springel 2003). However, we note that here we
adopted a less extreme wind model than Springel \& Hernquist (2003b),
with only half of the available SN-II powering the wind, not nearly
all of it, as they assumed. If we had also adopted such stronger
winds, the residual overcooling in our clusters would have been
reduced.

This therefore indicates that additional feedback processes may be at
work in the highly overdense environments of clusters and groups of
galaxies, and that perhaps additional sources of energy beyond SN--II
are involved.  Based on semi--analytical modelling of galaxy
formation, Menci \& Cavaliere (2000) and Bower et al. (2001) argued
that the feedback energy available from SN-II should actually be
enough to heat the ICM to the desired level, the problem however is
that this requires a very high, possibly unrealistic, efficiency for
the thermalization of this energy. Of course, given a fixed energy
budget available for heating, one can invoke `optimal' ways of
releasing it to the ICM to maximize its impact. For instance, one can
postulate that energy feedback targets just those particles which are
about to undergo cooling (e.g., Kay et al. 2003) or which have long
enough cooling time (e.g., Marri \& White 2003). Although such ad-hoc
schemes are often explored in feedback recipes, one would prefer if
they could be shown to arise as a natural consequence of a physically
self--consistent model.

Other sources of energy appear therefore as increasingly attractive
possibilities, for example SN-Ia. The energetics of SN-Ia is usually
considered to be subdominant, otherwise they would produce too much
Iron and overpollute the ICM (e.g., Renzini 1997, 2003; Pipino et
al. 2002). On the other hand, because the progenitors of SN-Ia have a
much longer life--time than those of SN-II, the corresponding energy
is released more gradually into a medium which is already heated by
the shorter--lived SN-II. This heated gas has hence already a longer
cooling time, making SN-Ia potentially a much more efficient heating
source than SN-II. However, a proper implementation in simulations
requires that the assumption of instantaneous recycling is dropped
(e.g., Lia, Portinari \& Carraro 2002; Valdarnini 2003; Kay et
al. 2003; Kawata \& Gibson 2003; Tornatore et al. 2003, in
preparation).

A further source of energy feedback, which is not included in our
simulation, is represented by AGN.  The energy output of AGN can be
extremely large, of the order $2 \times 10^{62} (M_{\rm BH}/10^9
M_{\odot})$ erg, where $M_{\rm BH}$ is mass of the central supermassive
black--hole. Theoretical calculations show that a fractional coupling
of the energy released by the AGN with the surrounding ICM at the
level of $f \approx 0.01$ would be sufficent to account for the
$L_X$--$T$ relations of groups (Cavaliere et al. 2002). While a number
of hydrodynamical cosmological codes now include a treatment for star
formation and SN feedback, none of them includes yet a
self--consistent treatment of energy release from AGN. As a first
approximation to their effect, the corresponding feedback energy could
simply be added to the energy budget provided by SN, which should be
adequate as long as the nuclear activity follows the star formation
within the hosting galaxies (e.g., Franceschini et al. 1999). However,
dynamically consistent models of AGN within cosmological simulations
of structure formation are clearly needed to understand their effects
in detail.

Perhaps the most puzzling discrepancy with observations concerns the
temperature profiles. Cooling causes a lack of pressure support in the
cluster center, causing gas to flow in from outer regions, being
heated by adiabatic compression (e.g., Tornatore et al. 2003). As a
result, the temperature actually increases in cooling regions, causing
steeply increasing temperature profiles. This picture is quite
discrepant compared to the standard cooling-flow model, where a
population of gas particles at very low temperature should be detected
(see Fabian 1994, for a review). Even more importantly, it also
conflicts with the observational picture emerging from Chandra and
XMM--Newton observations: the ICM in cluster central regions has
temperatures between 1/2 and 1/4 of the virial temperature, with no
signature for the presence of colder gas (e.g., Kaastra et al. 2001;
Peterson et al. 2001; B\"ohringer et al. 2002). It is this gas, which
is not allowed to cool below that temperature and drop out of the hot
phase, that causes the smooth decrease of the central temperature
profiles.  In fact, observations point now towards a quite small rate
of cooling in central cluster regions, with mass--deposition rates
reduced by a factor 5--10 with respect to pre--Chandra/XMM
observations (e.g., David et al. 2001; Blanton, Sarazin \& McNamara
2003).

One interesting mechanism that has been suggested to regulate gas
cooling in the central regions is thermal conduction. This process may
heat the gas in the central regions by generating a heat current from
external layers, thereby offsetting cooling losses such that the gas
can remain in the diffuse phase at a relatively low temperature for a
long time. Analytical computations have shown that this mechanisms,
possibly in combination with internal heating from AGN, can actually
reproduce realistic temperature profiles (e.g., Zakamska \& Narayan
2002; Ruszkowski \& Begelman 2002) with values of the effective
conductivity of about 1/3 of the Spitzer value (Spitzer 1962), while
also regulating gas cooling (e.g., Voigt et al. 2002). However, given
the ubiquitousness of magnetic fields in clusters, there is
considerable uncertainty whether the effective conductivity can really
reach sizable fractions of the Spitzer value (Brighenti \& Mathews
2003). Also, small--scale temperature variations of the ICM are now
constraining the thermal conductivity to be as small as one--tenth of
the Spitzer value (e.g., Markevitch et al. 2003), unless special
magnetic field configurations produce thermally isolated regions.

In the light of the above discussion, it is clear that the challenge
for numerical simulations of cluster formation has shifted from
problems related to resolution and dynamic range to those concerned
with the proper treatment of the complex physical processes which
determine the thermal state of cosmic baryons. The simulation
presented here demonstrates that code efficiency and super--computing
capabilities make it possible to describe cosmic structure formation
over a fairly large dynamic range. With the ever growing
super--computing power, the real challenge for numerical cosmology in
the coming years will be to construct algorithms that more faithfully
incorporate all those astrophysical processes that are relevant to
understand the properties of galaxies and clusters of galaxies.

\section*{Acknowledgments.} The simulation has been realized using the
IBM-SP4 machine at the ``Centro Interuniversitario del Nord-Est per il
Calcolo Elettronico'' (CINECA, Bologna), with CPU time assigned thanks
to an INAF--CINECA grant. We are grateful to Claudio Gheller for his
continuous assistance during the run. 
We are grateful to the referee, Trevor Ponman, for insightful comments
which improved the presentation of the result.
We acknowledge useful discussions with Stefano Ettori, Pasquale
Mazzotta and Mark Voit. We also thank Sabrina De Grandi, Alexis
Finoguenov, Yasushi Ikebe and Trevor Ponman for having provided the
files of the data shown in Figs. \ref{fi:mt}, \ref{fi:tproj},
\ref{fi:xtf} and \ref{fi:sthr}, respectively. This work has been
partially supported by INFN (Grant PD-51), by MIUR (Grant
2001, prot. 2001028932, ``Clusters and groups of galaxies: the
interplay of dark and baryonic matter''), by ASI and by the NATO
Collaborative Linkage Grant PST.CLG.976902.

\end{document}